\documentclass[%
 aip,
 amsmath,amssymb,
 reprint,%
]{revtex4-1}

\usepackage{graphicx}% Include figure files
\usepackage{dcolumn}% Align table columns on decimal point
\usepackage{bm}% bold math
\usepackage[normalem]{ulem}
\usepackage[utf8]{inputenc}
\usepackage[T1]{fontenc}
\usepackage{mathptmx}
\usepackage{etoolbox}
\usepackage{physics}
\usepackage{amsmath}
\usepackage{xcolor}
\makeatletter
\def\@email#1#2{%
 \endgroup
 \patchcmd{\titleblock@produce}
  {\frontmatter@RRAPformat}
  {\frontmatter@RRAPformat{\produce@RRAP{*#1\href{mailto:#2}{#2}}}\frontmatter@RRAPformat}
  {}{}
}%
\makeatother
\begin{document}

\preprint{AIP/123-QED}

\title[Surface charges  in a Rydberg atom-nanowaveguide hybrid quantum system]{Surface charges in a Rydberg atom-nanowaveguide hybrid quantum system}
% Force line breaks with \\
\author{Aswathy Raj}%
\thanks{These authors contributed equally to this work.}
\affiliation{Light-Matter Interactions for Quantum Technologies Unit, Okinawa Institute of Science and Technology Graduate University, Onna, Okinawa
904-0495, Japan}
\email{aswathy@oist.jp}
\author{Anna Kortel}
\thanks{These authors contributed equally to this work.}
\affiliation{Light-Matter Interactions for Quantum Technologies Unit, Okinawa Institute of Science and Technology Graduate University, Onna, Okinawa
904-0495, Japan}
\author{Krishna Jadeja}
\affiliation{Light-Matter Interactions for Quantum Technologies Unit, Okinawa Institute of Science and Technology Graduate University, Onna, Okinawa
904-0495, Japan}
\author{Dylan J. Brown}
\affiliation{Light-Matter Interactions for Quantum Technologies Unit, Okinawa Institute of Science and Technology Graduate University, Onna, Okinawa
904-0495, Japan}
\affiliation{Centre for Cold Matter, Blackett Laboratory, Imperial College London, London SW7 2AZ, United Kingdom}
\author{Alexey Vylegzhanin}
\affiliation{Light-Matter Interactions for Quantum Technologies Unit, Okinawa Institute of Science and Technology Graduate University, Onna, Okinawa 904-0495, Japan}
\author{Robert L{\"o}w}
\affiliation{5th Institute of Physics, Center for Integrated Quantum Science and Technology (IQST), University of Stuttgart, Pfaffenwaldring 57, 70569 Stuttgart, Germany}
\author{S\'ile {Nic Chormaic}}
\affiliation{Light-Matter Interactions for Quantum Technologies Unit, Okinawa Institute of Science and Technology Graduate University, Onna, Okinawa
904-0495, Japan}
\email{sile.nicchormaic@oist.jp}

\date{\today}% It is always \today, today,
             %  but any date may be explicitly specified

\begin{abstract}
Hybrid quantum platforms based on highly excited Rydberg atoms coupled to nanophotonic devices offer a promising route toward scalable quantum networks and integrated quantum technologies. However, the close proximity of Rydberg atoms to dielectric nanostructures makes these systems particularly susceptible to uncontrolled surface electric fields that can lead to a degradation of the excitation process. Here, we experimentally investigate Rydberg excitation of laser-cooled $^{87}$Rb atoms via the evanescent field of an optical nanofiber in the presence of fiber-guided red- and blue-detuned light fields as used to trap ground state atoms in fiber-based dipole traps. We observe a time evolution of the Rydberg excitation spectrum when both the dipole trapping fields are on and the additional spectral features that appear can be suppressed by applying an external oscillating electric field to the system, strongly indicating that surface charge accumulation is responsible for the observed spectral feature.  The experimental results are reproduced qualitatively by a model that incorporates DC energy level shifts arising from electric fields generated by charges deposited on the nanofiber surface.  We  identify Rydberg–ground state collisional ionization, which is enhanced by the dipole trapping fields, as the dominant mechanism for charge generation. These results provide new insight into charge dynamics at dielectric nanophotonic interfaces and establish practical guidelines for mitigating surface charge-induced electric fields in fiber-integrated Rydberg quantum systems.

\end{abstract}

\maketitle

\section{\label{sec:1}Introduction}

Hybrid quantum platforms combine the advantages of two or more physical systems, thereby enabling enhanced functionality and expanding the range of potential applications beyond what can be achieved using a single system alone. Coupling atoms to nanophotonic devices provides a route to scalable, integrated quantum technologies, due to the strong light field confinement and efficient photon collection\cite{Kimble2013, thompson2013coupling, dhordjevic2021entanglement, skljarow2022purcell, wave_guide} and optical nanofibers (ONFs) are currently one of the most mature interfaces\cite{LeKien2004, NicChormaic:06, Warken:07,  Aoki2015, self-org2016, le2018enhancement, corzo2019waveguide, Lamsal:19, nayak2019real, Ray_2020Light, McDonnell2022subradiantedge, berroir2025ultralow,413r-dn5p, bhavya2026interfacing, pache2026lambda, takahata2026fiber}. ONFs provide light-atom interaction lengths over several millimeters compared with the micrometer scale achievable for free-space beams of similar diameter, and they are directly compatible with optical fiber networks.  Scalability can be achieved by connecting multiple atom-ONF nodes in series\cite{Memory2015, sunami2025scalable}, providing efficient, low-loss quantum information transfer over long distances \cite{wave_guide}.  

Combining ONFs with Rydberg atoms is particularly interesting because the strong dipole-dipole interactions and large polarizabilities of Rydberg states can significantly extend the capabilities of ONF-based quantum devices\cite{saffman2016quantum, adams2019rydberg, stourm2020spontaneous, rajasree2020generation, stourm2023interaction, vylegzhanin2023excitation}.  Notably, Rydberg atoms are extremely sensitive to electric fields due to their large polarizability{\cite{abel2011electrometry, holloway2022rydberg, li2023super, ocola2024control}, and this feature can be exploited for quantum sensing and quantum metrology\cite{sensing1999, sedlacek2012microwave, ding2022enhanced, zhang2026microwave}. This sensitivity results in spectral splitting, shifts, and broadening, allowing Rydberg states to be used for precise electric field sensing \cite{grabowski2006high, viteau2011rydberg}. 

The same exceptional electric field sensitivity that makes Rydberg atoms so attractive for quantum technologies can also be a serious limitation, as they are highly susceptible to uncontrolled electric fields near dielectric surfaces\cite{negative_electron_affinity, epple2017effect}.  Understanding the nature and origin of these fields is, therefore, essential for realizing robust Rydberg atom-ONF quantum interfaces.  Many proposed Rydberg atom-nanophotonic platforms require atoms to be trapped in well-defined arrays near the nanophotonic surface \cite{rajasree2020generation,  vylegzhanin2023excitation, PhysRevLett.132.113601, Vylegzhanin_2025c}.  For ONF-based systems, this can be achieved either by using fiber-guided two-color optical dipole traps \cite{LeKien2004, vetsch2010optical, gupta2022machine, 89l1-dys4}  for the ground state atoms or by the combination of optical tweezers with ONF-guided light fields \cite{nayak2019real, vylegzhanin2026light, takahata2026fiber}. In general, fiber-guided dipole traps are especially attractive since they provide a compact trapping geometry that does not require alignment of additional free-space beams.   However, the close proximity of the Rydberg atoms to the dielectric ONF introduces additional challenges, since ionization processes\cite{nedeljkovic2005ionization, neufeld2011probing, kohlhoff2016interaction} could lead to charge accumulation on the ONF surface, producing stray electric fields that limit the excitation of Rydberg states. Nevertheless, Rydberg excitation of magneto-optically trapped $^{87}$Rb atoms up to  $n = 68$ via an ONF has been reported\cite{vylegzhanin2023excitation}. Recent work has shown that surface charges can substantially modify the trapping potential near an ONF and can be used in conjunction with Casimir-Polder interactions and blue-detuned light to increase the storage time beyond that achievable in  conventional ONF-based dipole traps\cite{pennetta2025hybrid}.  This hybrid platform also pushes the trap position to more than $600$~nm from the fiber surface, an aspect that could prove highly advantageous for Rydberg atom studies near ONFs, and further emphasizes the importance of understanding the origin of such charges and how their accumulation on dielectric nanophotonic interfaces can be controlled.

Here, we investigate the influence of fiber-guided dipole trapping beams on Rydberg excitation near an ONF using cold $^{87}$Rb atoms. We limit our discussion to $nD_{5/2}$ Rydberg states with $n= 27 - 40$.  The larger dipole matrix elements of the $D_{5/2}$ states result in a higher excitation probability and it has previously been shown that these states are less affected by the ONF\cite{vylegzhanin2023excitation}.  We observe a time-dependent evolution of the Rydberg excitation spectrum under specific ONF-guided light configurations. We model the spectral features by considering the effect of a stray DC electric field and identify the mechanism most likely responsible for generating the field. Our results provide new insight into Rydberg atom excitations close to an ONF in the presence of high-intensity evanescent fields and establish practical guidelines for mitigating electric field-induced charge effects in a Rydberg atom-ONF platform. 

\section{\label{sec:2}Experiment}

\subsection{\label{sec:2a} Experimental Setup}
The setup consisted of a silica ONF with a diameter of~$\sim$370~nm overlapped with a cloud of $^{87}$Rb atoms in a standard magneto-optical trap (MOT). A schematic of the experiment is shown in Fig. \ref{fig:Exp_details}(a) and further details can be found in our earlier works \cite{rajasree2020generation, vylegzhanin2023excitation}.  The atom cloud was imaged by collecting the MOT fluorescence in free-space. The fluorescence signal was then equally divided between an electron multiplying charge-coupled device (EMCCD, ANDOR Technology, Luca$^{EM}$ R, DL-604M-OEM) and a photomultiplier tube (PMT, Hamamatsu R636-10). 

Rydberg excitation was via a two-photon process, illustrated in Fig. \ref{fig:Exp_details}(b), where the first photon was provided by the 780~nm cooling beams that address the $5S_{1/2} \xrightarrow{}  5P_{3/2}$ transition and the second photon, which addresses the~$5P_{3/2} \xrightarrow{} nD_{5/2}$ transition, was from 480~nm ONF-guided light. The 480~nm light was from a Toptica SHG Pro laser that was stabilized to the desired frequency using an electromagnetically induced transparency (EIT) technique in a $^{87}$Rb-enriched vapor cell \cite{rajasree20211}.  To scan the 480~nm frequency, we used a fiber electro-optic modulator (EOM, NIR-NPX800, Photline Technologies) to change the 780~nm probe frequency. The 480~nm frequency followed correspondingly to maintain the EIT condition. Taking into account the Doppler shift, a 1~MHz shift in the EOM frequency corresponds to a 
$\Delta_{480} = -(780/480) \times \Delta_{780} \approx$ 1.6~MHz shift in the 480~nm frequency. 

\begin{figure}[htb]
\centering
\includegraphics[height=10cm]{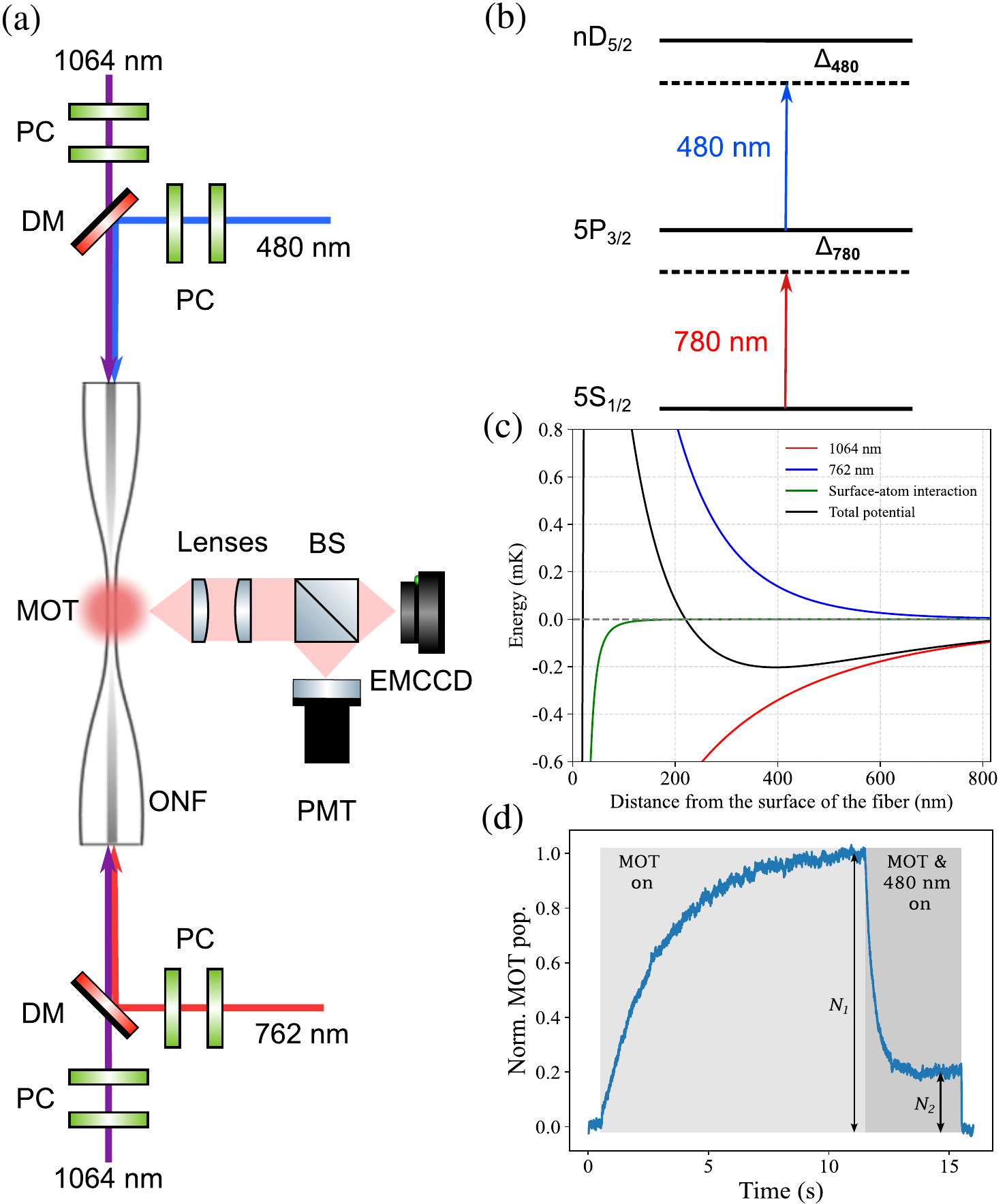}
\caption{(a) Schematic of the experimental setup. The polarization control (PC) consists of two quarter-wave plates. The beams are combined using dichroic mirrors (DM). The cloud of atoms is aligned with the waist of the ONF and its fluorescence is monitored by a PMT and an EMCCD. (b) Diagram of the Rydberg excitation scheme, with detunings $\Delta_{780}$ and $\Delta_{480}$. (c) Numerical simulation of the ONF-based dipole trap potential, showing attractive (red) and repulsive (blue) potentials, and the total potential (black) including atom-surface interactions (van der Waals and Casimir-Polder) (green). The trap position is 406~nm away from the fiber surface and the trap depth is 0.17~mK. (d) Sample MOT population curve at two-photon resonance for the $35D_{5/2}$~state. $N_{1}$ is the MOT saturation value with no 480~nm fiber-guided light and $N_{2}$~is the MOT equilibrium after switching on the 480~nm light.}
\label{fig:Exp_details}
\end{figure} 

We introduced a two-color dipole trap at the interface of the ONF, similar to that described in Gupta et al.\cite{gupta2022machine} Counterpropagating 1064~nm beams of $\sim$2~mW each and a single 762~nm beam of $\sim$1~mW were guided through the ONF. The 1064~nm light was red-detuned from the $D_2$ line of $^{87}$Rb, yielding an attractive potential for the ground state, while the 762~nm light provided a repulsive potential.  The resultant trapping potential had a minimum at around 400~nm from the ONF surface (Fig. \ref{fig:Exp_details}(c)), with the red- and blue-detuned trapping beams being collinearly polarized. The Rydberg excitation beam (480~nm, 30~$\mu$W) and the dipole trapping beams were  linearly polarized along the same axis using a set of quarter-wave plates to control the polarization at the waist of the ONF\cite{tkachenko2019polarisation}. 

\subsection{\label{sec:2b}Measurement Procedure}

The main goal of this study was to investigate the influence of the high-intensity evanescent fields generated by the fiber-guided  dipole trapping beams on the Rydberg excitation process.  We considered various combinations of the 1064~nm and 762~nm dipole trap beams propagating through the ONF, together with the 480~nm Rydberg excitation light (see Section \ref{sec:3a} for further details).  To isolate the influence of the fiber-guided trapping beams on the Rydberg excitation spectra while avoiding the added complexity of trapped-atom dynamics, all measurements were performed on atoms in the MOT while the dipole trapping beams were on but without deliberately loading atoms into the dipole traps.   

The experimental sequence was as follows: initially, atoms were loaded into the MOT for 11~seconds,  to reach a steady-state MOT population, $N_{1}$, see Fig. \ref{fig:Exp_details}(d). We normalized the MOT population in all measurements such that $N_1=1$. Then the fiber-guided 480~nm Rydberg excitation laser was switched on for 4 seconds.  The atoms near the ONF interacted with the 780~nm cooling beams and the 480~nm evanescent field, with some undergoing Rydberg excitation. Excitation to the Rydberg state satisfied the two-photon resonance condition when $\Delta_{480} = -\Delta_{780}$. Excited atoms were lost from the MOT, leading to a decrease in the fluorescence signal. We kept the 480~nm light on until the normalized atom population in the MOT reached a new equilibrium, $N_{2} \leq 1$.  It is important to note that the MOT population decrease of more than $50\%$ is not simply due to Rydberg atoms leaving the MOT, but also to collisions of Rydberg state atoms and ground state atoms~\cite{gallagher1988rydberg}, Langevin interactions of ionized Rydberg atoms, Rydberg atom inner ionic core and ground state atoms~\cite{harter2012single}, and the interaction between ground state atoms and Rydberg atom outer electrons.

To record the Rydberg excitation spectrum, the dependence of $N_{2}$ on the 480~nm laser detuning was obtained by scanning $\Delta_{480}$. For each value of $\Delta_{480}$, the measurement was repeated four times and the corresponding $N_{2}$ values were averaged, to yield $\bar{N}_{2}$, see Fig. \ref{fig:split spectrum}.

\subsection{\label{sec:2c}Results}

\subsubsection{\label{sec:3a}Rydberg Excitation Spectra for Dipole Trap Beam Configurations} 

We considered three different dipole trapping beam configurations in the ONF: (i) a unidirectional 1~mW 762~nm beam, (ii) counterpropagating beams of 1064~nm, each with a power of 2~mW, and (iii) the combination of both (i) and (ii), i.e., the two-color dipole trapping beams. Because the coupling efficiencies and splice losses differ for each wavelength, the exact powers at the ONF waist cannot be measured directly. We therefore quote the powers measured at the output pigtail, noting that the power at the waist could be higher due to losses in the uptaper. The transmissions through to the ONF output pigtail with respect to the input are for 762~nm $\sim$40\%,  1064~nm $\sim$55\%, and 480~nm $\sim$10\%. For all guided wavelengths except 480~nm, the ONF is single-mode;  for 480~nm light we assume that power coupling to higher order modes was negligible. 

The Rydberg excitation spectra are shown in Fig. \ref{fig:split spectrum}(a). Each spectrum was acquired over approximately 1 hour over this range of $\Delta_{480}$. The beams for each configuration remained on during the measurement,  the 480~nm light was on for 4~seconds, and each data point was averaged over 4 measurements. Configurations (i) and (ii) yielded very similar spectra (blue and red curves), exhibiting the Autler-Townes splitting reported in our earlier work\cite{vylegzhanin2023excitation}.  In contrast, config.~(iii) displayed broadening and additional spectral features.

Next, we deliberately exposed the system to continuous 480 nm light, tuned to $\Delta_{480}=+20$~MHz for 30 minutes with the MOT on and acquired a spectrum  following the same procedure as described for Fig. \ref{fig:split spectrum}(a). The results are shown in Fig. \ref{fig:split spectrum}(b). The spectra for configs. (i) and (ii) are largely unchanged. However, config.~(iii) shows that the broadening in Fig. \ref{fig:split spectrum}(a) develops into diverse spectral features.  We infer that the combined trapping fields influence the Rydberg excitation process when the 480~nm light is on for longer times, yielding  a qualitatively different  spectrum. 

\begin{figure}
\centering
\includegraphics[height=10.5cm]{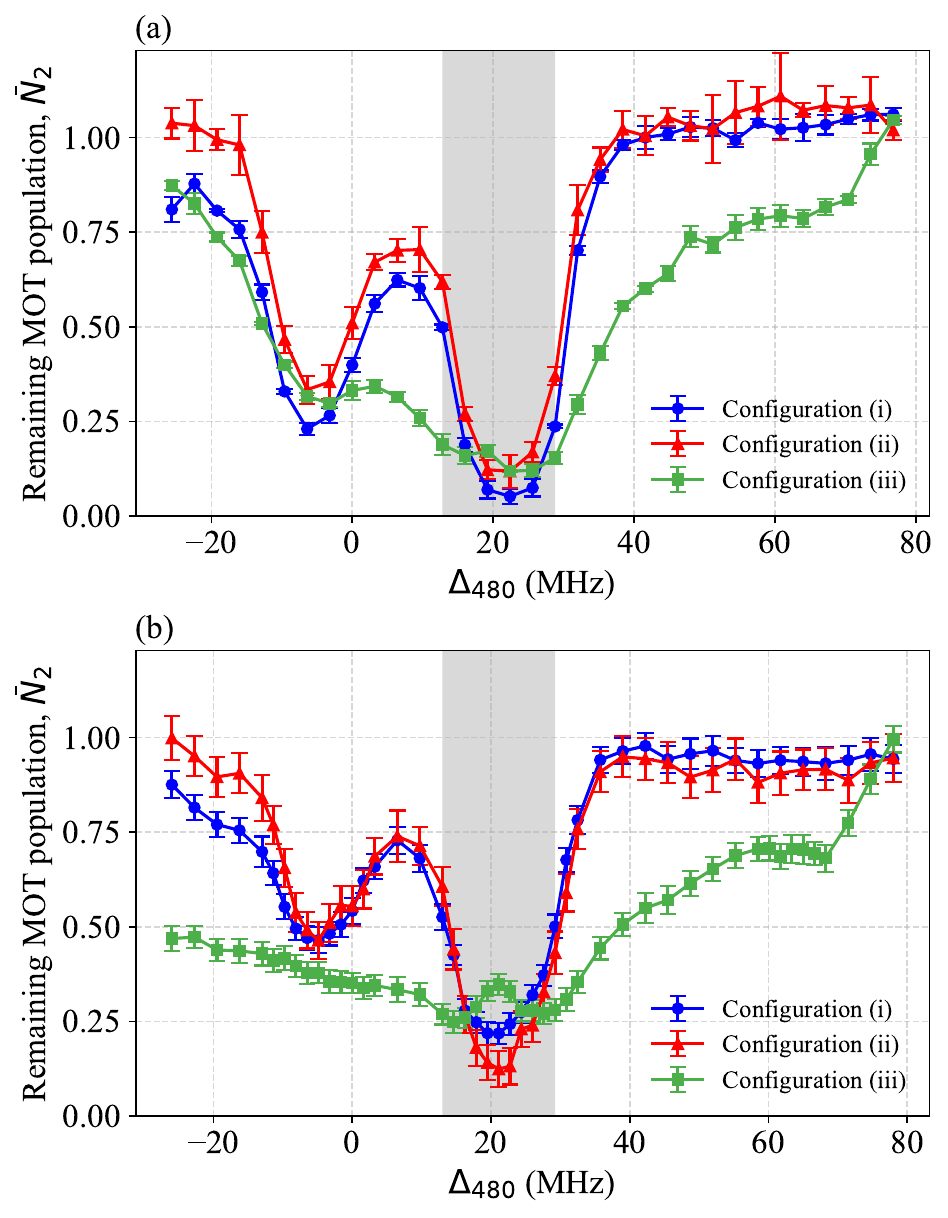}
\caption{Rydberg excitation spectra for $35D_{5/2}$ for three trapping beam configurations in the ONF: (i) a unidirectional 762~nm beam (blue curve), (ii) counterpropagating beams of 1064~nm (red curve), and (iii) the combination of both wavelengths (green curve). For each configuration, the corresponding beams were left on during the measurement. (a)~Initial spectrum for each beam configuration.  (b) Spectra after the system was exposed to continuous 480 nm light, tuned to $\Delta_{480}=+20$~MHz for 30 minutes with the MOT on.  In both (a) and (b) each data point corresponds to 4 seconds of 480~nm exposure, averaged over 4 measurements.} 
\label{fig:split spectrum}
\end{figure}

To better understand the origin of the spectral features observed in Fig.~\ref{fig:split spectrum}, we considered the time evolution of the spectrum while scanning the 480~nm detuning from 0 to 40~MHz to reduce spectrum measurement times.  We began when the system had not been exposed to 480 nm light for several hours. We measured the steady-state MOT fluorescence without any 480 nm light in the ONF, $N'_1$, and normalized this to 1. We detected the fluorescence from the MOT via the PMT for 5 seconds, averaged over the full PMT signal, and normalized this to $N'_1$, giving the value for $\bar{N'_2}$, see Fig. \ref{fig:time_evolution}. The initial measurement was performed as soon as the MOT reached an equilibrium population after the 480~nm light was switched on.  It took approximately 10 – 15 minutes to take each subsequent spectrum and the 480~nm light was kept on until all spectra were acquired. As can be seen from the figure, as the exposure time increases, the deformation of the spectrum becomes more pronounced.  

\begin{figure}
\centering
\includegraphics[height=6.5cm]{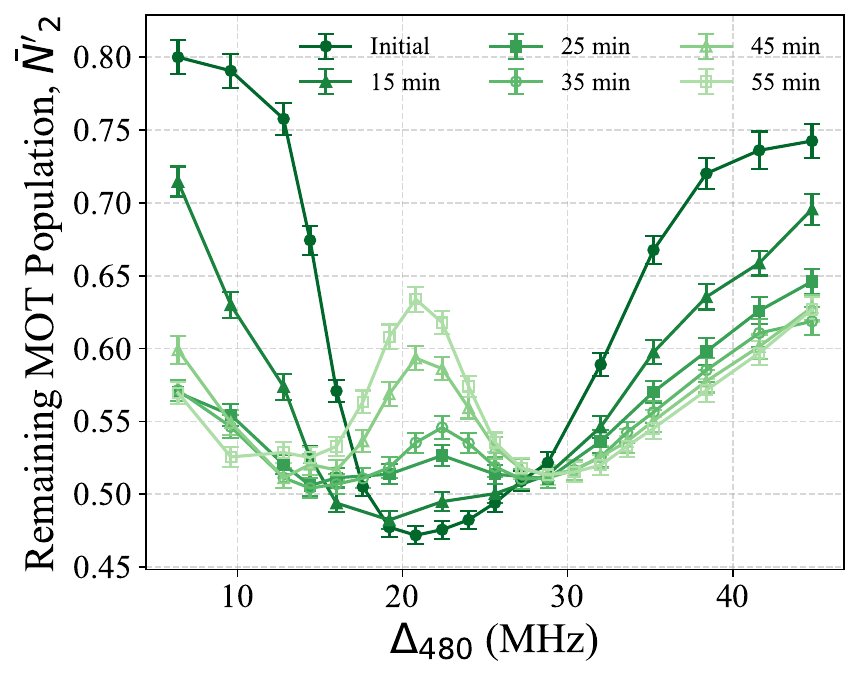}
\caption{Time evolution of the Rydberg excitation spectrum for $35D_{5/2}$  for config.~(iii) with 480~nm light continuously on. The legend gives the cumulative 480~nm exposure time, beginning from when the 480~nm light was initially switched on. Each data point corresponds to 5~seconds of 480~nm exposure and averaged over the 5~second PMT signal.}
\label{fig:time_evolution}
\end{figure}

\subsubsection{\label{sec:3b}Origin of the Spectral Features}

Since it is known that Rydberg atoms near dielectric surfaces are susceptible to ionization\cite{nedeljkovic2005ionization, neufeld2011probing, kohlhoff2016interaction}, we developed a method to prevent charge deposition on the ONF to see if it influences the Rydberg excitation process. We applied an external AC electric field to the setup using a pair of ring electrodes (inner diameter: 3 cm, outer diameter: 4 cm, thickness: 2 mm) set up along the axis of one of the MOT beams (see Fig.~\ref{fig:ring electrode}(a)). These produced a homogeneous electric field with $\sim$0.7~V/cm amplitude at the nanofiber region. The voltage applied was alternating at 100~kHz to prevent either electrons or Rb ions from depositing on the vacuum chamber’s viewports.

\begin{figure}
\centering
\includegraphics[height=15.0cm]{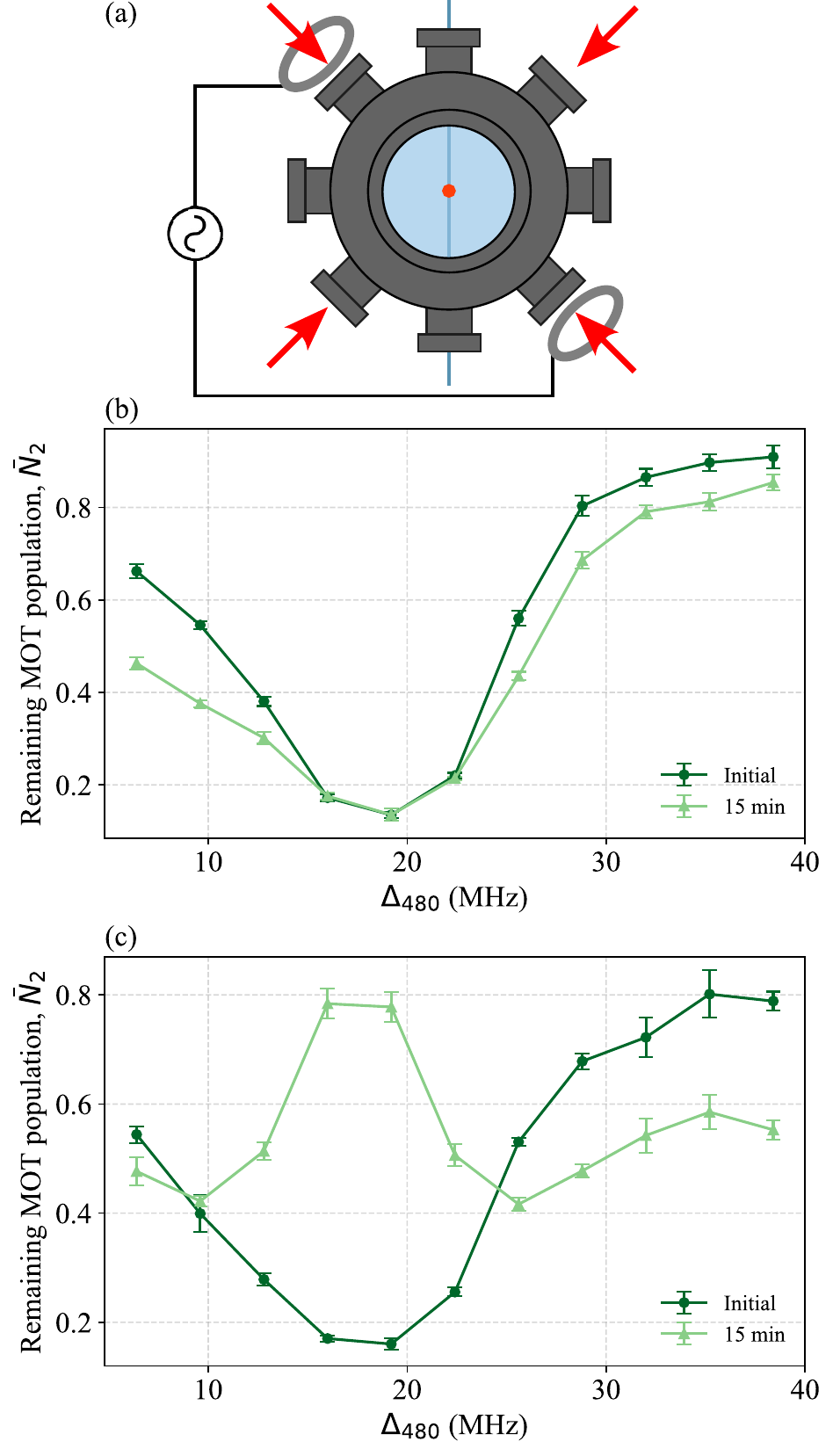}
\caption{(a) Experimental setup schematic showing the orientation of electrodes added to the system to produce an AC field, where the vertical line depicts the fiber, red arrows depict MOT beams and the red circle depicts the MOT. (b) and (c) Rydberg excitation spectra for $40D_{5/2}$ for config.~(iii). Initial spectrum (dark green) and spectrum after 15 minutes of 480~nm exposure (light green) (b) with and (c)  without the applied AC   field. Each data point corresponds to 4~seconds of 480~nm exposure, averaged over 4~measurements}
\label{fig:ring electrode}
\end{figure}

We considered the $40D_{5/2}$ state as it is more sensitive to DC fields than $35D_{5/2}$\cite{bai2019single, meyer2020assessment}.  The measurement procedure followed that in Fig. \ref{fig:split spectrum}, with the region of interest being the shaded area. An initial spectrum was first acquired where there had been no exposure to 480~nm light in config.~(iii) for several hours prior to the measurement, see Fig. \ref{fig:ring electrode}(b) (dark green curve). We then exposed the system to 15 minutes of 480~nm light (at $+20$~MHz detuning), while simultaneously applying the external AC field. After the 15 minutes, we switched off the AC field to avoid spectral shifts, and then acquired another excitation spectrum, see Fig. \ref{fig:ring electrode}(b) (light green curve). We see no evidence of spectral deformation.   Next, we followed the exact same procedure, without the AC field during the 15 minutes of 480~nm exposure, see Fig. \ref{fig:ring electrode}(c). Here we see clear evidence of splitting appearing in the spectrum (light green curve).} 
 
The application of the AC field was shown to suppress spectral splitting even after up to 75 minutes of 480~nm light exposure, see Appendix~\ref{sec:Appendix_A}, Fig. \ref{fig:ring_electrodes_75min}. The oscillating electric field  causes the acceleration of any generated charges along the axis of the plates, preventing them from settling on the ONF.   With the change observed, it is reasonable to conclude that charge accumulation on the ONF  contributes to the time evolution of the spectra, as observed in Figs. \ref{fig:split spectrum} and \ref{fig:time_evolution}.

As electrons accelerate $\sim 1.6 \times 10^5$ times faster than Rb ions, electrons move sufficiently far away from the fiber within half an electric field oscillation cycle, resulting in the reduced probability of their deposition on the ONF compared to Rb ions. Although surface charges could, in principle, provide a mechanism for electromechanical actuation of the ONF using an external AC electric field\cite{Wuttke2013, Tebbenjohanns2023}, this effect was not explored in this work. In our system, the applied field amplitude, ONF geometry, and mechanical resonance frequencies differ from those in previous studies; in particular, the applied 100~kHz field is not resonant with the mechanical modes of the ONF and no measurable mechanical response is observed. We next develop a numerical model that incorporates DC electric field effects on the Rydberg states and compare numerical spectra with experimental results. 

\section{\label{sec:4}Discussion}

\subsection{\label{sec:4a}Numerical Model for the Rydberg Excitation Spectra}

To model the origin of the deformations in the Rydberg excitation spectrum for config.~(iii) (Figs. \ref{fig:split spectrum} and \ref{fig:time_evolution}), we consider a multilevel system consisting of the ground ($5S_{1/2}$), intermediate ($5P_{3/2}$), and Rydberg ($nD_{5/2}$) states, split into six $m_j$ levels (Appendix~\ref{sec:Appendix_B}, Fig. \ref{Rydberg_mj_level_only_diagram}). The energy levels are labeled from $\ket{1}$ to $\ket{8}$, where $\ket{1} \xrightarrow{} \ket{2}$ is the first photon transition and $\ket{2} \xrightarrow{} \ket{k}$ ($k=3,...,8$) is the second photon transition. The dynamics of this multilevel system can be solved numerically using the Lindblad master equation

\begin{equation}
    \frac{d\rho}{dt} = -\frac{i}{\hbar}[H,\rho] + \mathcal{L},
\label{eq:Lindblad_formula}
\end{equation}
where $\rho$ is the density matrix of the system, $H$ is the atom-light interaction Hamiltonian and $\mathcal{L}$ is the Lindblad operator (for details see Appendix \ref{sec:Appendix_B}). The Hamiltonian  is given by 
\begin{widetext}
\begin{equation}
H = \frac{\hbar}{2}
\begin{bmatrix}
    0 & \Omega_{12} & 0 & \cdots & 0 \\
    \Omega_{12} & -2 \Delta_{780} & \Omega_{23} & \cdots & \Omega_{28}  \\
    0 & \Omega_{23} & -2(\Delta_{780}+\Delta_{480}+\delta_{\text{DC},3}) & \cdots  & 0 \\
    \vdots & \vdots & \vdots & \ddots & \vdots \\
    0 & \Omega_{28} & 0 & \cdots  & -2(\Delta_{780}+\Delta_{480}+\delta_{\text{DC},8})
\end{bmatrix}.
\label{eq:Hamiltonian}
\end{equation} 
\end{widetext}
 
\noindent The off-diagonal elements in Equation (\ref{eq:Hamiltonian}) represent the Rabi frequencies for the transitions. $\Omega_{12}$ drives the excitation from $5S_{1/2} \xrightarrow{} 5P_{3/2}$, while $\Omega_{2k}$ drives the transition from  $5P_{3/2} 
\xrightarrow{}$~$35D_{5/2}$ $m_{j}$ states.

\begin{figure}
\centering
\includegraphics[height=9.5cm]{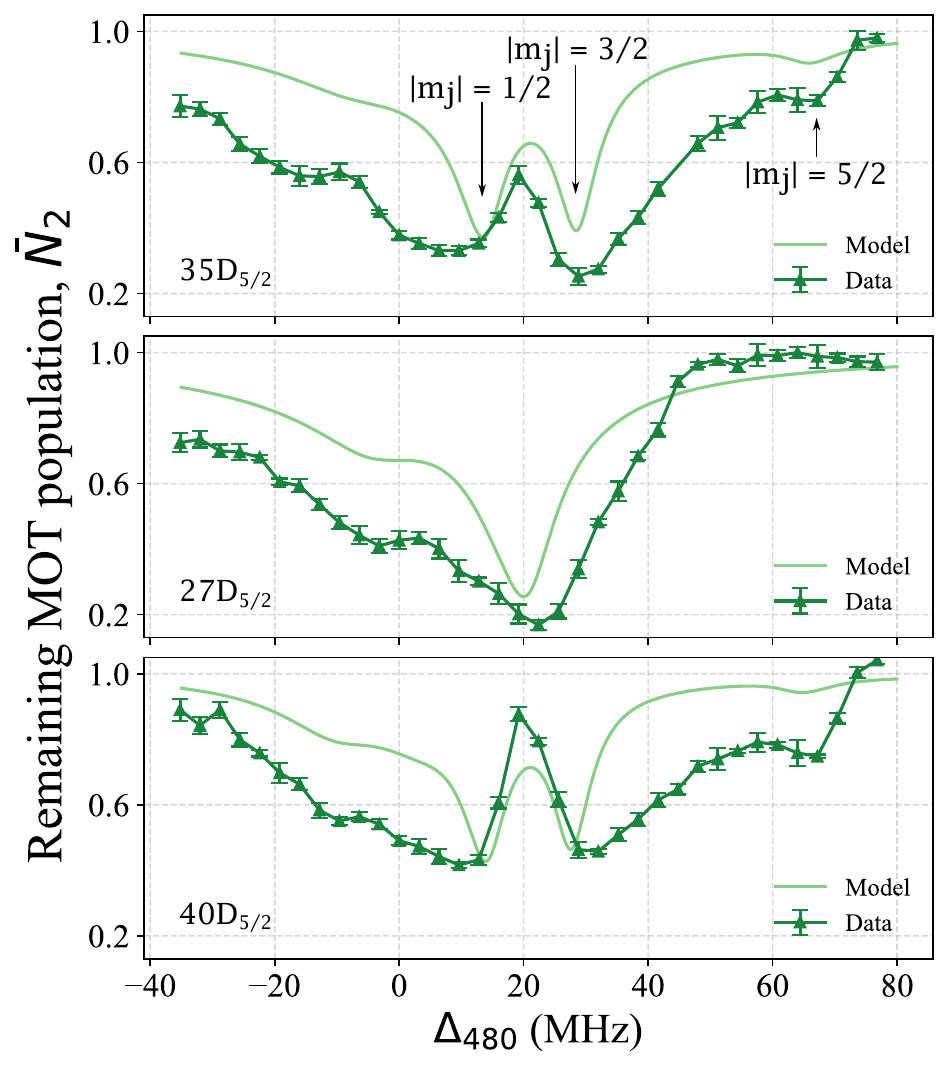}
\caption{Rydberg excitation spectra for $35D_{5/2}$, $27D_{5/2}$, and $40D_{5/2}$ for config.~(iii). Light green curve:  Numerical steady-state solution ($1-\rho_{rr}$) from Equation~(\ref{eq:Lindblad_formula}). Dark green curve:  Experimental data. For each configuration, the corresponding beams were left on during the measurement. Each data point corresponds to 4 seconds of 480~nm exposure, averaged over 4 measurements.}
\label{fig:modelling all}
\end{figure}

Since we were able to change the spectral features by minimizing charge accumulation on the ONF via an applied AC field, we include the effect of a stray DC electric field by introducing an additional spectral shift represented by $\delta_{\text{DC},k}$~ =  $-\frac{1}{2}\alpha^{k}_{\text{}}$E$^{2}_{\text{DC}}$ \cite{bai2019single}, where $\alpha^{k}_{\text{DC}}$ are the static electric field polarizabilities of each of the Rydberg $m_j$ states, and E$_{\text{DC}}$ denotes the DC electric field that is treated as a free parameter to match the model with the experimental data in Fig. \ref{fig:modelling all}. The DC polarizabilities of each of the Rydberg $m_{j}$ states were obtained from the Alkali-Rydberg Calculator (ARC) package \cite{vsibalic2017arc}. Since the DC polarizabilities of the ground and intermediate state are negligible compared to the Rydberg states, the effect of the static electric field on these states is also negligible.

Note that the AC light shifts due to the dipole trapping beams were not included in the model as no shift in the Rydberg excitation resonance was experimentally observed.

We solve Equation (\ref{eq:Lindblad_formula}) and obtain the population of atoms in the Rydberg $m_{j}$ levels, $\rho_{rr} = \sum_{k=3}^{8} \rho_{kk} $. We assume that the Rydberg excitation occurs at a distance of around 300-500~nm from the fiber surface.  Below 300~nm strong Casimir-Polder interactions restrict the excitations \cite{vylegzhanin2023excitation}, whereas beyond 500~nm the evanescent field decay limits Rydberg excitation through insufficient Rabi coupling. In the chosen range, the effects of van der Waals and Casimir-Polder interactions are negligible and are, therefore, not included. We sample 50~atoms along the axis of polarization of the 480~nm light field, which is set to be linear; however, due to the nature of light polarization in the ONF, there is a degree of ellipticity that must be considered\cite{le2004field}. We assume 90$\%$ of the Rydberg excitations occur with linearly polarized 480~nm light and the remaining 10$\%$  with circularly polarized light. 

From the experiment, the Rabi frequency of 780~nm is taken as 17~MHz and the detuning is -16~MHz. The DC polarizabilities for $35D_{5/2}$, $\abs{m_j}$ = (1/2, 3/2, 5/2) are (-1.8, 3.38, 16.77)~MHz~cm$^{2}$~V$^{-2}$ respectively. For the decay parameters we take $\Gamma_{12} = 2 \pi \times 6$~MHz and $\Gamma_{2k}$ =  $2 \pi \times 2.179$~kHz \cite{vsibalic2017arc}. For the dephasing parameters  we select  $\gamma_{2} = 2 \pi \times 3$~MHz and $\gamma_{k}$ =  2$\pi \times 1.5$~MHz, which take into account  collision-induced dephasing and laser linewidths of 780~nm and 480~nm, respectively.  

The results of our theoretical model, along with the experimental data for the $35D_{5/2}$ state, are shown in Fig. \ref{fig:modelling all}(top). From the model, the DC field is $\sim$2.45~V/cm at 300 nm and decreases to 2.35~V/cm at 500 nm from the ONF. Three resonances are observed corresponding to the $\abs{m_{j}}$ splitting of 1/2, 3/2 and 5/2. As our model does not account for all broadening mechanisms present in the system, such as Rydberg-Rydberg interactions, atom-ion or atom-electron collisions, Rydberg-ion or Rydberg-electron interactions, the result only matches the experimental data qualitatively. Nevertheless, a very good agreement in the overall shape is obtained.  This simplified model, along with the experimental data from Fig.~\ref{fig:ring electrode}, strongly indicate that a DC electric field is indeed responsible for the spectral deformations observed in the Rydberg spectra in Figs. \ref{fig:split spectrum} and \ref{fig:time_evolution}.

To identify the behavior of DC field splitting in other Rydberg states, we conducted equivalent measurements for the $27D_{5/2}$ and $40D_{5/2}$ states, see Fig. \ref{fig:modelling all}.  For $27D_{5/2}$ we observed broadening of the Rydberg spectrum rather than splitting, hence, no DC field was included in the model for this state. This is due to the lower sensitivity to DC fields of Rydberg states with lower principal quantum numbers\cite{bai2019single, meyer2020assessment}. In contrast, for the $40D_{5/2}$ state, the model matches experimental data best when we consider that excitation occurs in the range of 400 - 500~nm from the nanofiber surface. This yields a DC field of $\sim$1.5 V/cm at 400 nm which decreases to 1.4~V/cm at 500 nm. 

\subsection{\label{sec:4b}Numerical Model for the DC Field}

\begin{figure}[htb]
\centering
\includegraphics[height=6.5cm]{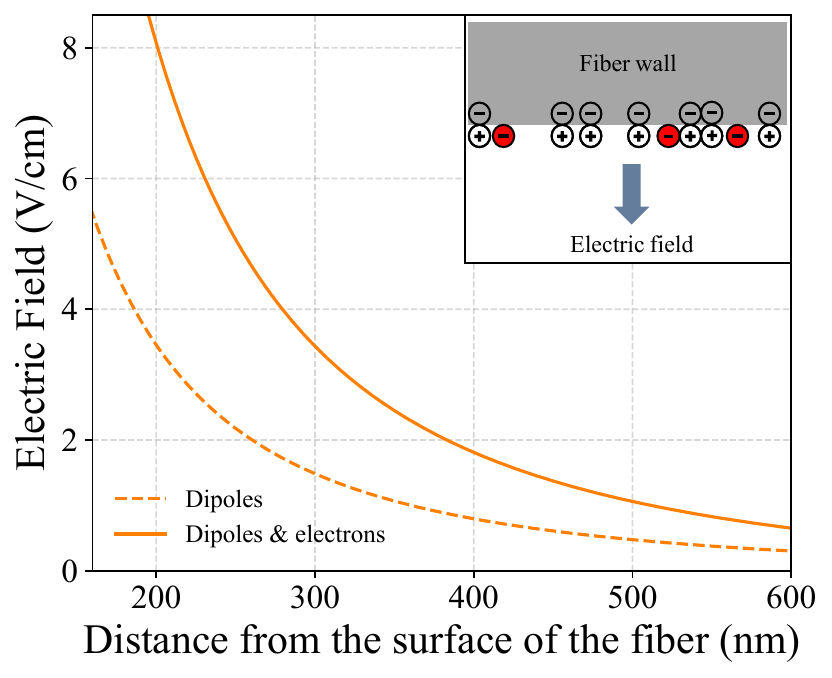}
\caption{Electric field as a function of the distance from the ONF surface, with an estimate of $5\times 10^5$~dipoles (dashed) plus 65~electrons (solid) stuck on the fiber. Inset: Schematic illustrating the radial DC electric field originating from dipoles (transparent circles) of Rb atoms adsorbed on the fiber surface. Red circles depict the random distribution of electrons on the ONF surface.}
\label{fig:lab2}
\end{figure}

Numerous studies have reported the presence of a stray electric field arising from dipoles generated by adsorbate atoms on dielectric surfaces \cite{negative_electron_affinity, hattermann2012detrimental, chan2014adsorbate, epple2017effect} and both our experimental data and theoretical model support this. As the Rb atoms are very close to the ONF, we can reasonably assume that there is a small background DC field present. Rubidium atoms are adsorbed on the ONF  due to the partial negative charge transfer to the oxygen in SiO$_2$. The adsorbed atoms create a dipole moment normal to the surface (Fig. \ref{fig:lab2} inset) and induce a negative electron affinity on the ONF.  This is favorable for the slow-moving electrons to bind to the surface \cite{negative_electron_affinity}. Here, we consider the possible contributions to the DC field from both Rb adsorbates and electrons deposited on the ONF. 

To estimate the number of dipoles contributing to the effective field, we consider the maximum DC field  ($\sim$1~V/cm at 350~nm from the fiber surface) in our model that will not induce any resolvable splitting in the experimental data for the $35D_{5/2}$ state. Similarly, for the $40D_{5/2}$ state, the maximum DC field that does not induce observable splitting is 0.8~V/cm at 400 nm. These fields were found using the model in Section \ref{sec:4a}, and compared to experimental data with no discernible splitting. The dipole contribution to the field is not tied to the creation of Rydberg atoms or ionization. We consider the dipole moment for Rb on the fiber surface to be $d_0 \approx 10~\text{Debye}$ per adsorbed atom \cite{mcguirk2004alkali,tauschinsky2010spatially,sedlacek2016electric}. By randomly distributing dipoles on the ONF surface, we find that $5\times 10^5$~dipoles produce the aforementioned field (Fig. \ref{fig:lab2} dashed curve). 

The electric field estimated in Section \ref{sec:4a} that induces splitting in the $35D_{5/2}$ ($40D_{5/2}$) state is within the range of 2.45~V/cm to 2.35~V/cm (1.5~V/cm to 1.4~V/cm) at distances of 300 – 500 nm (400 – 500 nm) from the ONF surface. We model the number of electrons randomly distributed on the fiber amongst the estimated number of dipoles such that they match the DC electric field ranges obtained in Section \ref{sec:4a} and find that $\sim$65 electrons together with the dipoles produce this field, see Fig. \ref{fig:lab2} (solid curve). The presence of the electrons in the dipole field
may eventually prevent further charge deposition, leading to a saturated
spectrum over time with the 480~nm laser exposure and in config.~(iii), see Appendix~\ref{sec:Appendix_C}, Fig. \ref{fig:saturation_plots}.    

The DC field model is consistent with the fields obtained from the numerical model in Section \ref{sec:4a}, supporting the hypothesis that the DC field experienced by the Rydberg atoms comes from the combined electric fields of Rb dipoles and electrons randomly distributed on the ONF surface. According to the plot shown in Fig. \ref{fig:lab2} (solid curve), excitation to the $35D_{5/2}$ state may occur closer to the ONF compared to the $40D_{5/2}$ state. This is likely due to the greater DC shifts experienced by the $40D_{5/2}$ state, given its higher sensitivity to DC electric fields, thereby limiting its excitation nearer the fiber.

\subsection{\label{sec:4c}Ionization Mechanism}

We identified that, during Rydberg excitation, both dipole trapping beams i.e., config.~(iii), were necessary to create the DC field, and we next considered the effect of the specific wavelength combination used. The 762~nm laser was replaced  by another blue-detuned laser at 755~nm. The power of the 755~nm light was set to maintain the same trapping potential as  before. With this new wavelength choice, termed config.~(iv), we observed a similar splitting for $35D_{5/2}$ to that observed with config.~(iii), see Appendix~\ref{sec:Appendix_D}, Fig.~\ref{fig:755_added}. We conclude that the splitting is not dependent on the wavelength of the dipole trapping beams, but rather the total dipole potential created by them. 

We initially observed the DC field splitting when the polarizations of the dipole trapping beams and the Rydberg excitation beam were collinear,  resulting in a maximum trapping potential and maximum overlap between all three evanescent fields (480 nm, 762 nm, and 1064 nm). To determine the influence of polarization on the excitation spectra, we  investigated the importance of the overlap by examining the deformation of the spectrum when varying the relative polarizations of the 480~nm Rydberg excitation laser and the trapping beams.  

\begin{figure}[htb]
\includegraphics[height=10.5cm]{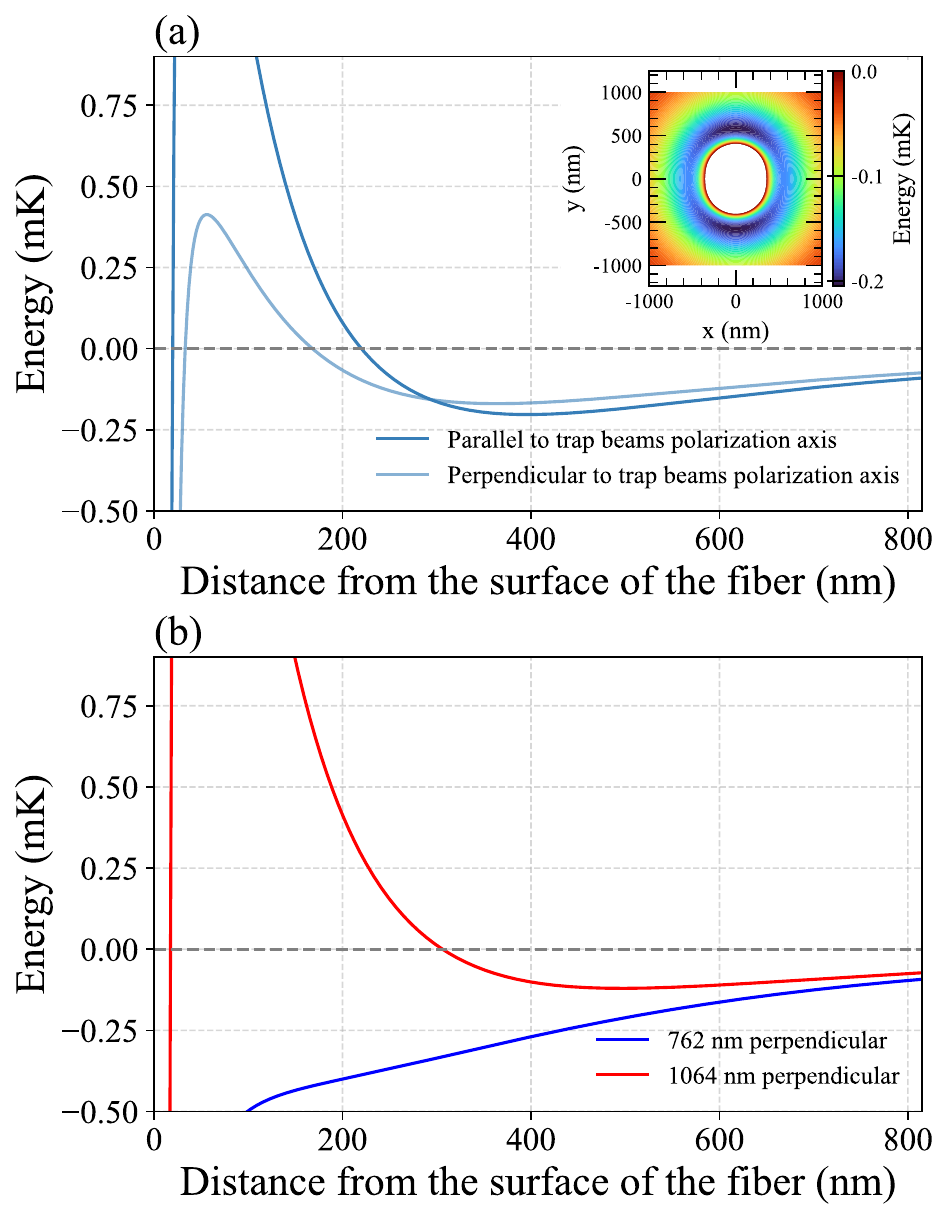}
\caption{(a) The total trapping potential along the axis of the 480~nm polarization direction, in the case where the polarizations of the trapping beams are collinear to each other. Dark blue curve: the potential when the 480~nm polarization direction is collinear with those of the trapping beams. Light blue curve: the potential when the trapping beam polarizations are perpendicular to that of the 480~nm light. Inset: 2D contour plot of the potential from the trapping beams in the $xy$ plane (i.e., normal to the axis of the ONF). The trapping beams are quasi-linearly polarized along the $y$-axis. (b) The trapping potential along the 480 nm polarization direction for the cases where either the polarization of the 762~nm beam (blue curve) or the 1064~nm  beam (red curve) is perpendicular to the other beams.}
    \label{fig:Potentials}
\end{figure}

First, we considered the polarization of the 480~nm beam aligned perpendicular to that of the trapping beams instead of parallel. The resulting spectrum was broadened, with less pronounced splitting compared to the collinear polarization case (Appendix~\ref{sec:Appendix_D}, Fig.~\ref{fig:480nm_pol_paralell_perp} (light blue)).  In Fig.~\ref{fig:Potentials}(a), we plot the total radial potential along the direction of the 480~nm polarization axis.  We see that, when this axis is perpendicular to that of the dipole trapping beams, the potential has a lower depth and is no longer confined in the azimuthal direction around the fiber (Fig. \ref{fig:Potentials}(a) inset), meaning there is less of an effect on the atomic density in this region. 

Next, we considered the perpendicular orientation of the linearly polarized trapping beams relative to each other (Fig. \ref{fig:Potentials}(b)) and its effect on the overall potential. When the blue-detuned 762~nm polarization is kept parallel to the 480~nm polarization, but the 1064~nm red-detuned polarization is perpendicular, there is still a radial potential minimum along the 480~nm polarization axis, see Fig.~\ref{fig:Potentials}(b) (red curve). With this configuration, we could still see some broadening, though less pronounced splitting of the excitation spectrum (Appendix~\ref{sec:Appendix_D}, Fig.~\ref{fig:1064_762_perp} (red curve)). However, when the 762~nm beam polarization was perpendicular to the 480~nm and 1064~nm polarizations, there is no stable trap minimum along the 480~nm polarization axis (Fig.~\ref{fig:Potentials}(b) (blue curve)) and there is no observed splitting of the excitation spectrum (Appendix~\ref{sec:Appendix_D}, Fig.~\ref{fig:1064_762_perp} (blue curve)). These results confirm that, by ensuring the greatest overlap between the Rydberg excitation field and that of the dipole trapping beams, the effect on the Rydberg excitation spectrum is maximized. 

Since the effective radius, $a_{eff}$, of the Rydberg atoms is large ($27D_{5/2}$, $a_{eff}= 35$~nm; $35D_{5/2}$, $a_{eff}= 60$~nm; $40D_{5/2}$, $a_{eff}= 79$~nm)\cite{vsibalic2017arc}, the probability of collisional ionization is quite high.  While we have identified that the trapping potential probably induces ionization, this may be due to its role in increasing the density of ground state atoms near the ONF, thereby effectively increasing Rydberg-ground state collisions during the Rydberg excitation process. We do not consider Rydberg-Rydberg collisions, as the potential from the dipole trapping beams is repulsive to Rydberg states. This reduces the number of Rydberg atoms in the vicinity of the fiber. 

To estimate the number of charges produced by Rydberg-ground state atom collisions, we calculate the ionization rate using the equation of the rate of collision $\Gamma_{\text{coll}}$ as
\begin{equation}
    \Gamma_{\text{coll}} = \mathcal{N}_{g}\cdot \sigma_{\text{coll}}\cdot\bar{v},
\label{eq:collision rate}
\end{equation}
where $\mathcal{N}_{g}$ is the ground state atom density. $\sigma_{\text{coll}}$ is the collisional cross-section between Rydberg and ground state atoms, which can be approximated as the geometric cross-section of the Rydberg atom with $\sigma_{\mathrm{coll}} = \pi \left(n^{*2} a_0\right)^2$, where $a_{0}$ is the Bohr radius and $n^{*}$ is the effective quantum number considering the quantum defect. $\bar{v}$ is the mean velocity of the atoms, calculated from their temperature of 200~$\mu$K, and is 0.22~m/s. For the case of $40D_{5/2}$, the ionization cross-section is calculated to be $3.53\times10^{-15}\mathrm{m}^{2}$, where we assume the ionization cross-section to be approximately 0.18$\sigma_{\mathrm{coll}}$ \cite{weller2016charge,weller2019interplay,Weller2019}, as not all collisions lead to ionization. For an atomic density in the MOT of around $2\times10^{10}~\mathrm{cm}^{-3}$, we obtain an ionization rate of 16~s$^{-1}$. A 15-minute 480~nm light exposure time (average time needed for the spectral deformations to emerge) can result in about 14,000 charges. From a solid angle estimation, the probability of generated electrons approaching the ONF is around 1 to 2$\%$ and is sufficient to produce the fields estimated in Section \ref{sec:4b}. 

We also verified the dependence of the charge creation on the 480 nm laser power. We show that a lower power of 480~nm light leads to the production of fewer Rydberg atoms, limiting the number of Rydberg-ground state atom collisions. This leads to a decrease in ionization rate, and subsequently, the charge in the system (see details in Appendix \ref{sec:Appendix_D}, Fig. \ref{fig:480nm_power_variation}).

We further verified the density-dependent effects in the $40D_{5/2}$ spectrum by comparing two different MOT densities, $2\times10^{9}~\mathrm{cm}^{-3}$ and $20\times10^{9}~\mathrm{cm}^{-3}$. We observed that different MOT densities exhibited similar DC field splitting, with differences in excitation probability. For $40D_{5/2}$, a higher density for a given charge production time leads to a lower excitation probability, indicating that the Rydberg excitation is happening further from the fiber. As mentioned in Section \ref{sec:4b}, a lower excitation probability indicates a higher electric field from the fiber, supporting that an increase in density will increase the ionization rate.

\section{\label{sec:5}Conclusion}

This study identifies experimental conditions under which charges can be generated in an ONF-Rydberg hybrid system, the mechanism by which this may occur, and a method for the prevention of charge deposition on the ONF. We observed DC field splitting of the $35D_{5/2}$ and $40D_{5/2}$ Rydberg excitation spectra in the presence of dipole trapping beams guided through the ONF. A numerical model was developed by solving the Lindblad master equation of a multilevel system. 

The DC fields estimated from the model in Section \ref{sec:4a}  are in the range of 2.45~V/cm to 2.35~V/cm at distances of 300 nm – 500 nm from the ONF surface for the split $35D_{5/2}$ state and in the range of 1.5~V/cm to 1.4~V/cm at 400~nm – 500~nm for the $40D_{5/2}$ state. The combined experimental and numerical results strongly support the conclusion that these DC fields arise from a combination of Rb dipoles and electrons randomly distributed on the ONF surface.  By applying an external electric field, we inhibited the spectral splitting, indicating that preventing charge accumulation on the ONF suppresses the observed spectral deformations. In addition, we investigated the role of the dipole trapping potential in increasing Rydberg-ground state atom collisions as a result of an increased density of the ground state atoms near the ONF. This, in turn,  may lead to a higher collisional ionization rate, which we propose is the primary mechanism for charge production. 

The identification of a likely charge generation mechanism in this work provides new insight into how surface charges may arise in optical nanofiber systems, complementing a recent demonstration that such charges can be exploited to realize hybrid nanophotonic traps\cite{pennetta2025hybrid}. Surface charging must, therefore, be regarded as an important design consideration for future nanophotonic atom interfaces, particularly those involving Rydberg atoms. The suppression strategy demonstrated here provides a practical route toward stable fiber-integrated Rydberg trapping and waveguide-QED experiments. As nanophotonic platforms evolve toward trapped arrays of Rydberg atoms and waveguide-mediated interactions, understanding the mechanisms responsible for charge accumulation—and developing practical strategies to suppress it—will be essential for realizing robust integrated quantum devices.

\section{\label{sec:5}Dedication}
    We are pleased to contribute to this special issue celebrating the 60th birthday of Ernst Rasel. Beyond his pioneering scientific achievements, Ernst has been a valued colleague, mentor, and friend to many in the cold atom and quantum optics communities. Some of the authors have known Ernst since the very early stages of his career as a PhD student in Innsbruck, and it has been a pleasure to witness both his scientific accomplishments and his support of the wider community over the years. We thank him for his inspiration, generosity, and continued friendship and leadership in the field, and look forward to his future contributions to the field.

\begin{acknowledgments}
The authors would like to thank Z.~Shahrabifarahani, S.~Abdrakhmanov, R.~Kumar, A.~Konovalov, K.~Rachek, R.K.~Gupta, A.~Pendse, and H.~Takahashi for useful discussions. We are grateful for the support provided by the Engineering Section and the Scientific Computing \&
Data Analysis Section of OIST. This work was funded by the Okinawa Institute of Science and Technology Graduate University (OIST), the Japan Society for the Promotion of Science (JSPS) KAKENHI Grant No. 24K08289 and Invitational Fellowships for Research in Japan (Short-term) Fellowship ID: S25153, the Japan Science and Technology Agency (JST) as part of Adopting Sustainable Partnerships for Innovative Research Ecosystem (ASPIRE), Grant No. JPMJAP2511, and Japan’s Council for Science, Technology and Innovation (CSTI) under the Cross-Ministerial Strategic Innovation Promotion Program (SIP) for “Promoting the application of advanced quantum technology platforms to social issues” (Funding agency: QST, Grant No. JPJ012367).
\end{acknowledgments}

\section*{Author Declarations}
\subsection*{Conflict of interest} 
The authors have no conflicts to disclose.
\subsection*{Author Contributions} 
\textbf{Aswathy Raj:} Methodology; Investigation; Data Curation; Formal Analysis; Writing - original draft; Writing - review $\&$ editing. \textbf{Anna Kortel:} Methodology; Investigation;  Data Curation; Formal Analysis; Writing - original draft; Writing - review $\&$ editing. \textbf{Krishna Jadeja:} Methodology; Investigation; Data Curation; Formal Analysis; Visualization; Writing - original draft; Writing - review $\&$ editing. \textbf{Dylan J. Brown:} Conceptualization, Methodology, Validation; Writing - review $\&$ editing. \textbf{Alexey Vylegzhanin:} Methodology, Writing - review $\&$ editing. \textbf{Robert L{\"o}w:} Conceptualization; Methodology; Validation; Funding Acquisition; Writing - review $\&$ editing. \textbf{Síle {Nic Chormaic}:} Supervision; Conceptualization; Methodology; Validation; Funding Acquisition; Writing - review $\&$ editing.  

\section*{Data Availability Statement}
Data available on reasonable request from the authors.

\appendix{\label{sec:Appendix}}
\section{Origin of the Spectral Features}{\label{sec:Appendix_A}}

\begin{figure}[h!]
\centering
 \includegraphics[height=6.5cm]{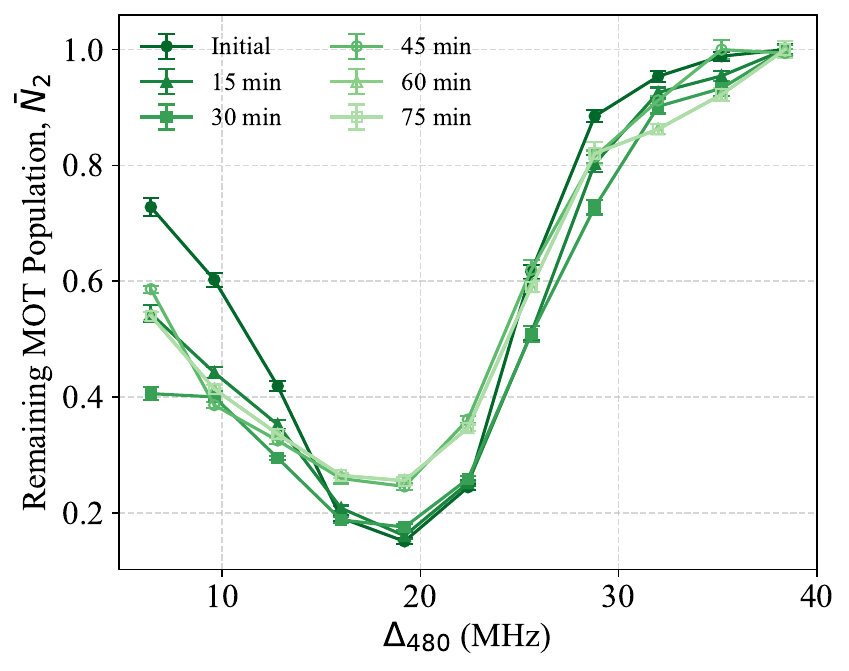}
\caption{Rydberg excitation spectra for $40D_{5/2}$.  The external AC electric field is applied while the ONF is exposed to 480~nm and config.~(iii) light in 15 minute intervals, with a measurement taken between each interval. The AC field was turned off while the measurements were taken. The labels depict the cumulative exposure time from the 15 minute intervals of 480~nm and config.~(iii) light exposure.}
\label{fig:ring_electrodes_75min}
\end{figure}

Here, we present additional data supporting Section \ref{sec:3b}, demonstrating the suppression of the spectral deformations for dipole trap config.~(iii) and a cumulative 75~minute exposure to 480~nm light. The total exposure time was divided into 15 minute intervals, with the spectrum taken after each interval. A slight broadening of the spectrum is observed with increasing 480~nm exposure; however, no splitting is observed, see Fig. \ref{fig:ring_electrodes_75min}   

\section{Numerical Model for Rydberg Excitation Spectrum}{\label{sec:Appendix_B}}

\begin{figure}[h!]
\centering
\includegraphics[height=6.5cm]{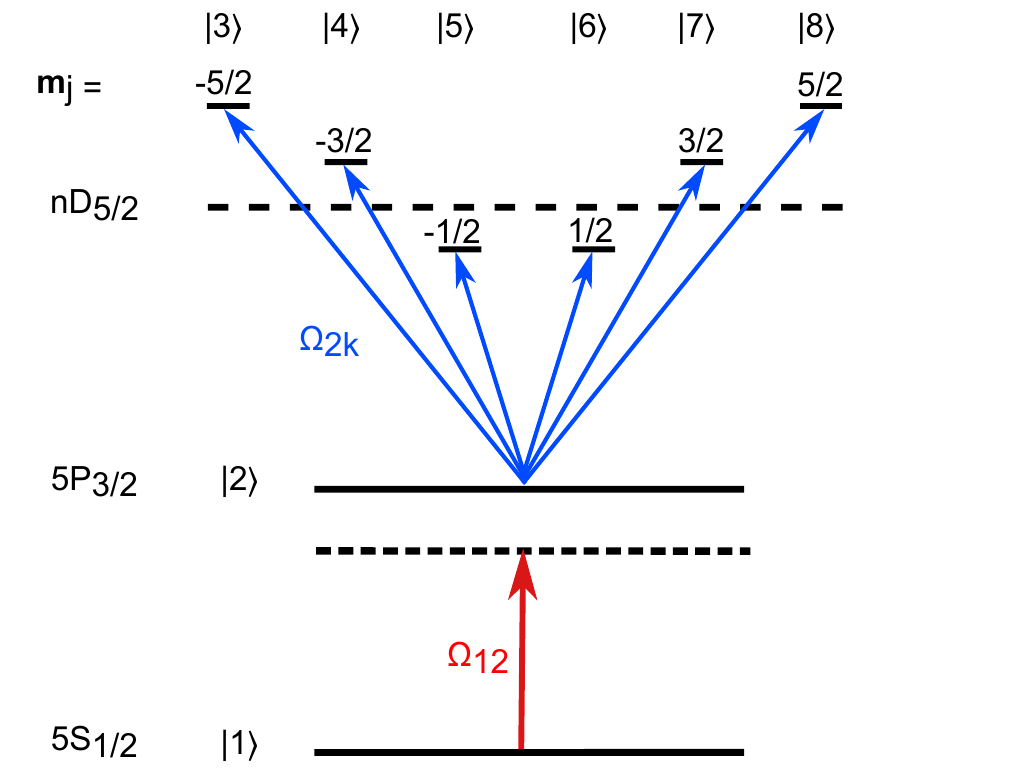}
\caption{Extended three-level energy level diagram of $^{87}$Rb including the Rydberg $m_j$ states.}
\label{Rydberg_mj_level_only_diagram}
\end{figure}

The Lindblad operator for the multi-level system used for the numerical model consists of two parts; one describes the spontaneous decay of the atom ($\mathcal{L}_{\text{decay}})$ and the other accounts for the dephasing ($\mathcal{L}_{\text{dephasing}})$, for example, from collisions between the atoms, and the finite linewidth of the excitation lasers. We have that\cite{lindblad1976generators}:
\vspace{0.5 cm}
\begin{equation}
\begin{aligned}
\mathcal{L}_{\text{decay}} & = \sum_{m_j}C_{ij}\rho C_{ij}^{\dagger} - \frac{1}{2}\left(C_{ij}^{\dagger}C_{ij}\rho + \rho C_{ij}^{\dagger}C_{ij}\right),\\
\mathcal{L}_{\text{dephasing}} & = \sum_{m_j}C_{i}\rho C_{i}^{\dagger} - \frac{1}{2}\left(C_{i}^{\dagger}C_{i}\rho + \rho C_{i}^{\dagger}C_{i}\right),
\end{aligned}
\label{eq:Lindblad2}
\end{equation}
where the decay operators are defined as
\begin{equation}
\begin{aligned}
    C_{ij} & = \sqrt{\Gamma_{j}}\ket{i}\bra{j},\\
    C_{i} & = \sqrt{\gamma_{i}}\ket{i}\bra{i}.
\end{aligned}
\label{eq:Lindblad3}
\end{equation}
Here, $\Gamma_{j}$ represents the spontaneous decay rate from state $\ket{j} \xrightarrow{}\ket{i}$. We assume that each of the Rydberg $m_{j}$ states decays back to the intermediate level, $\ket{2}$. The term $\gamma_{i}$ represents the dephasing rate of the states $\ket{i}$. We can now rewrite the Lindblad operator as   

\begin{widetext}
\begin{equation}
\mathcal{L} =
\begin{bmatrix}
    \Gamma_{21}\rho_{22} & -\frac{1}{2}\kappa_{2}\rho_{12} & -\frac{1}{2}\kappa_{3}\rho_{13} & \cdots & -\frac{1}{2}\kappa_{8}\rho_{18} \\
    -\frac{1}{2}\kappa_{2}\rho_{21} & -\Gamma_{21}\rho_{22} + \sum _k \Gamma_{k2}\rho_{kk} &  -\frac{1}{2}(\kappa_{2} + \kappa_{3})\rho_{23} & \cdots  & -\frac{1}{2}(\kappa_{2} + \kappa_{8})\rho_{28} \\
    -\frac{1}{2}\kappa_{3}\rho_{31} &  -\frac{1}{2}(\kappa_{3} + \kappa_{2})\rho_{32} & -\Gamma_{32}\rho_{33} & \cdots  & -\frac{1}{2}(\kappa_{3} + \kappa_{8})\rho_{38}  \\
    \vdots & \vdots  & \vdots  & \ddots & \vdots \\
    -\frac{1}{2}\kappa_{8}\rho_{81} &  -\frac{1}{2}(\kappa_{8} + \kappa_{2})\rho_{82}   & -\frac{1}{2}(\kappa_{8} + \kappa_{3})\rho_{83} & \cdots  & -\Gamma_{82}\rho_{88}
\end{bmatrix},
\end{equation} 
\end{widetext}
where
\begin{equation}
\begin{aligned}
\kappa_{2}& = \Gamma_{21}+\gamma_{2},\\
\kappa_{k}& = \Gamma_{k2}+\gamma_{k}.\\
\end{aligned}
\end{equation}

\section{Saturation Measurements}{\label{sec:Appendix_C}}

\begin{figure}[h!]
\centering
\includegraphics[height=9.5cm]{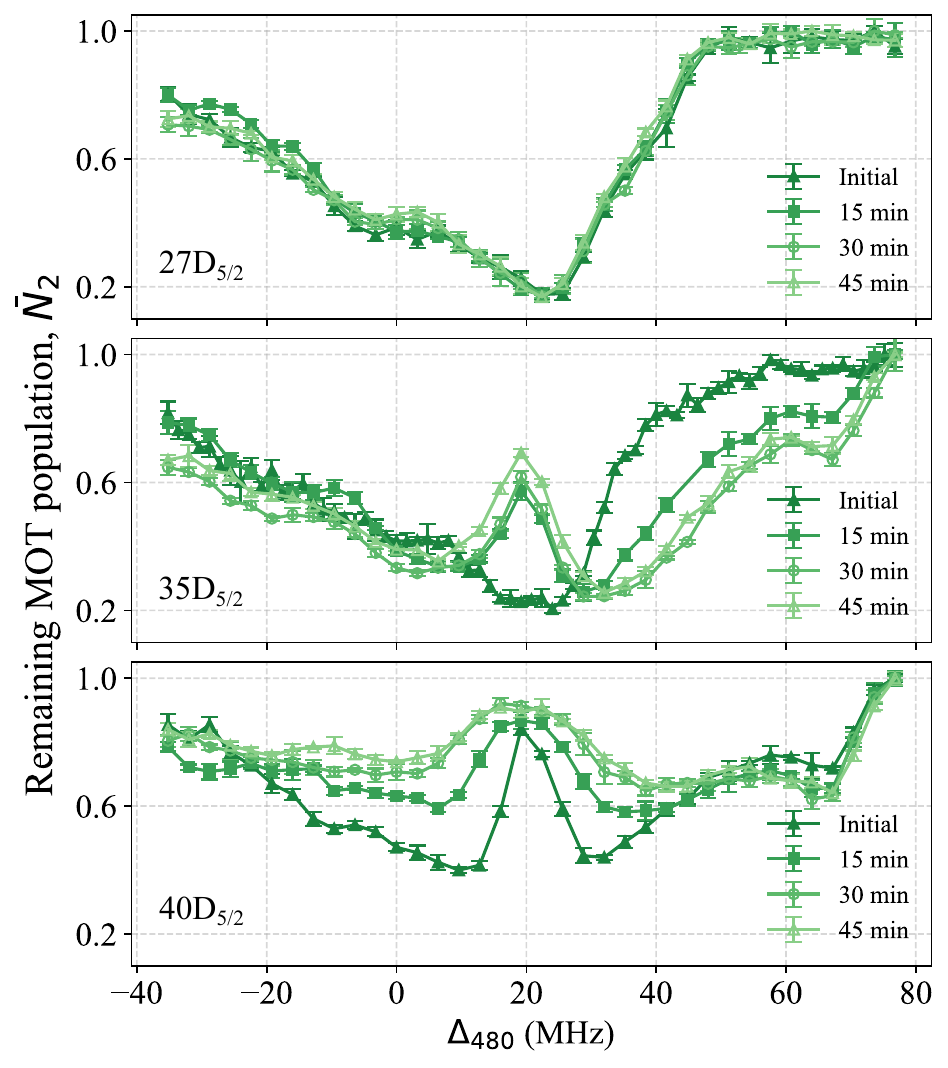}
\caption{Rydberg excitation spectra for $27D_{5/2}$, $35D_{5/2}$, and $40D_{5/2}$ for config.~(iii). The spectral features reached saturation with repeated measurements. The labels depict the cumulative exposure time from the 15 minute intervals of 480~nm and config.~(iii) light exposure.}
\label{fig:saturation_plots}
\end{figure}

To further validate the proposed explanation of the origin of the electric field in our system, we repeatedly performed measurements for 27$D_{5/2}$, 35$D_{5/2}$, and 40$D_{5/2}$ as outlined in Section \ref{sec:2b} until the spectral features no longer changed (see Fig. \ref{fig:saturation_plots}). This confirms that the presence of electrons may prevent further charge deposition on the ONF.

\vspace{1.2 cm}
\section{Ionization Mechanism}
\begin{figure}[h!]
\centering
\includegraphics[height=6.5cm]{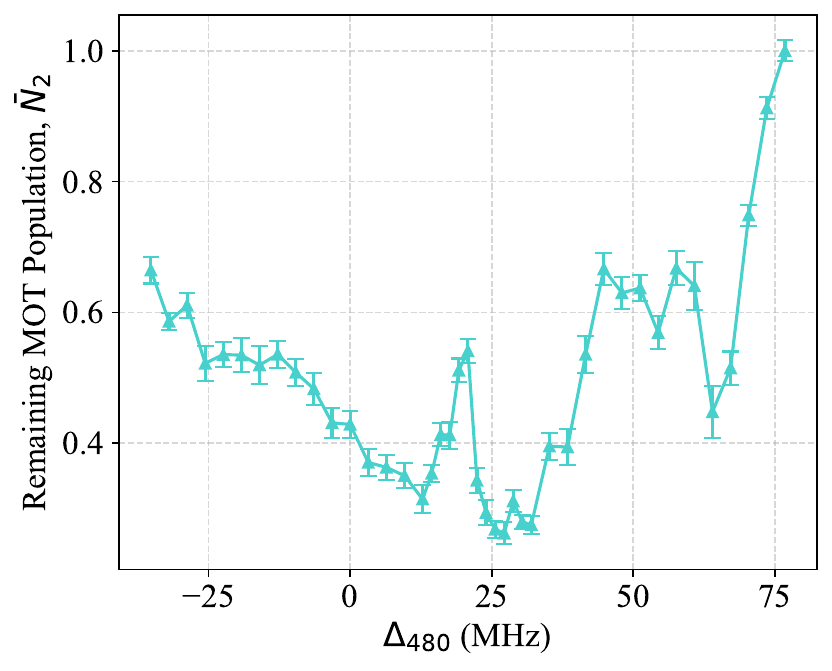}
\caption{Rydberg excitation spectrum for $35D_{5/2}$ for config.~(iv), consisting of 1064~nm counterpropagating beams and a single beam of 755~nm light in the ONF.}
\label{fig:755_added}
\end{figure}

{\label{sec:Appendix_D}}
To ascertain the mechanisms responsible for the increased ionization in the presence of the config.~(iii) beams, several tests were carried out, as discussed in Section \ref{sec:4c}. Figure~\ref{fig:755_added} shows the Rydberg excitation spectrum for the $35D_{5/2}$ state with beam config.~(iv), showing clear spectral splitting.

\begin{figure}[h!]
\centering
\includegraphics[height=6.5cm]{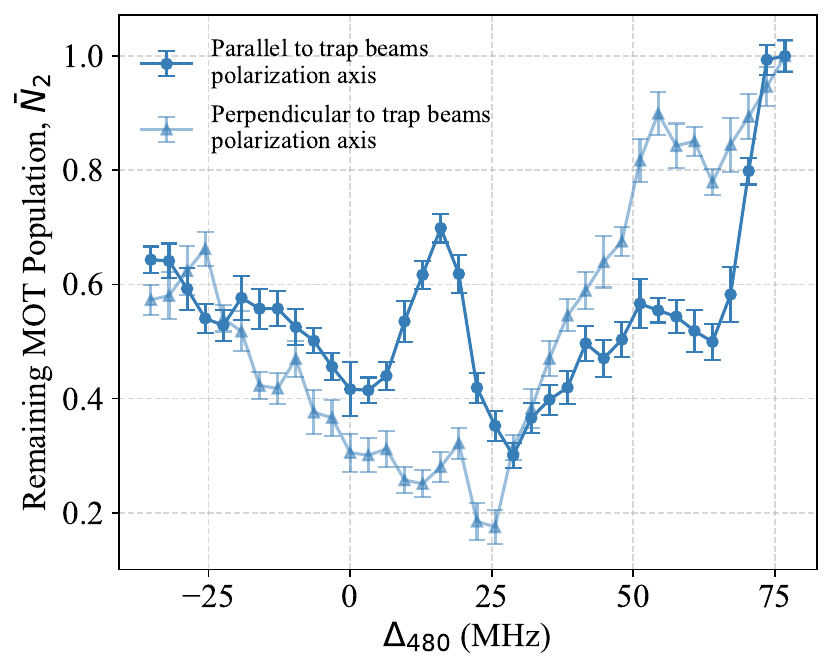}
\caption{Rydberg excitation spectra for $35D_{5/2}$ where the polarization direction of the 480~nm light is parallel (dark blue) and perpendicular (light blue) to the trapping beams polarization direction.}
\label{fig:480nm_pol_paralell_perp}
\end{figure}

\begin{figure}[h!]
\centering
\includegraphics[height=6.5cm]{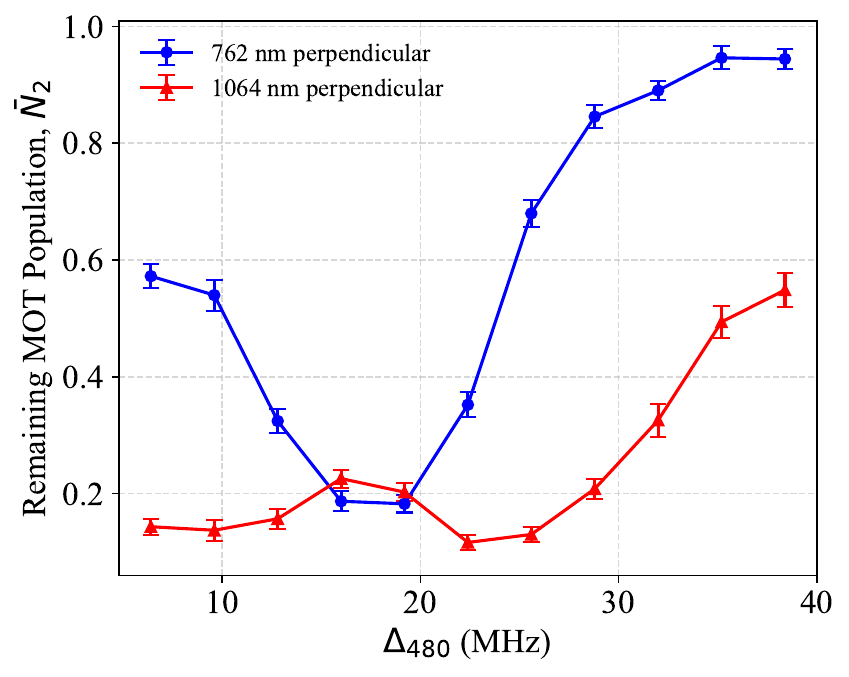}
\caption{Rydberg excitation spectra for $35D_{5/2}$, where the polarization direction of the 762~nm light is perpendicular to the 480~nm and 1064~nm polarization directions (blue) and the polarization direction of the 1064~nm light is perpendicular to the 480~nm and 762~nm polarization directions (red).}
\label{fig:1064_762_perp}
\end{figure}

Figure \ref{fig:480nm_pol_paralell_perp} shows the Rydberg excitation spectrum for the $35D_{5/2}$ state, with the 480~nm polarization direction parallel (dark blue curve) and perpendicular (light blue curve) to the trapping beams. When the 480~nm beam polarization direction is perpendicular to the trapping beams, the spectral features are less pronounced than for the parallel case. Considering the polarization direction of the config.~(iii) beams (Fig.~\ref{fig:1064_762_perp}), when the 762 nm beam's polarization direction is perpendicular to that of the 480~nm and 1064~nm light (blue curve), no spectral feature is observed. In contrast, when the 1064 nm polarization direction is perpendicular (red curve), the feature appears, but is less pronounced.

\begin{figure}[h!]
\centering
\includegraphics[height=6.5cm]{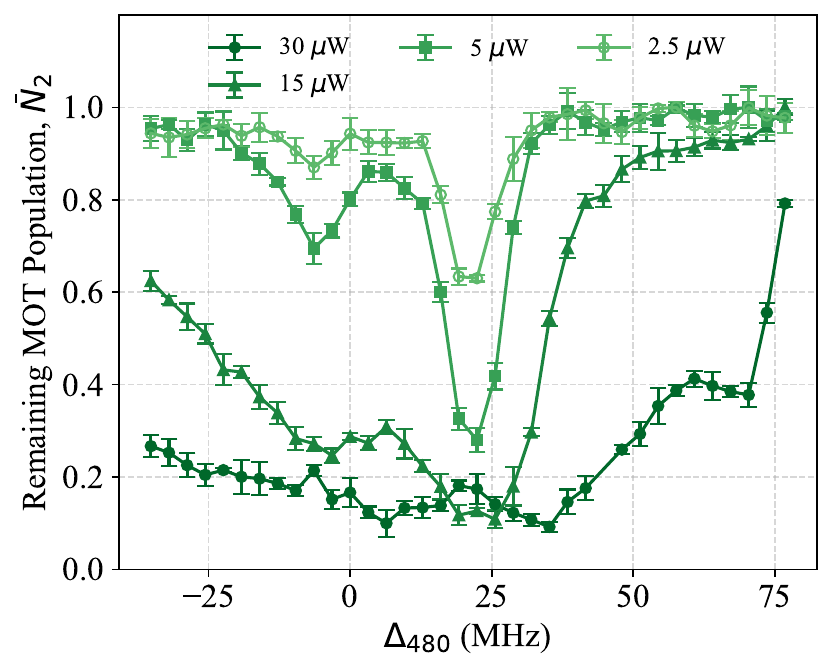}
\caption{Rydberg excitation spectra for $35D_{5/2}$ for config.~(iii) for different 480~nm Rydberg laser powers. The measurements were taken in order from high to low power.}
\label{fig:480nm_power_variation}
\end{figure}

Fig. \ref{fig:480nm_power_variation} shows the effect of 480~nm laser power on the appearance of Rydberg spectral splitting for $35D_{5/2}$ in config.~(iii). For 30~$\mu W$ of 480~nm light, we observed a very broadened spectrum with splitting of the Rydberg $m_j$ states (see Fig. \ref{fig:480nm_power_variation}, 30~$\mu W$ plot). Subsequent measurements were performed with 15~$\mu W$, 5~$\mu W$, and 2.5~$\mu W$ of 480~nm light at the ONF output. This clearly indicates that with lower 480~nm power, the Rydberg excitation spectrum in config.~(iii) shows the normal Autler-Townes splitting.

\clearpage
\section*{References}
\bibliography{References}

@PREAMBLE{
 "\providecommand{\noopsort}[1]{}" 
 # "\providecommand{\singleletter}[1]{#1}%" 
}

@article{Kimble2013,
  title = {Self-Organization of Atoms along a Nanophotonic Waveguide},
  author = {Chang, D. E. and Cirac, J. I. and Kimble, H. J.},
  journal = {Phys. Rev. Lett.},
  volume = {110},
  issue = {11},
  pages = {113606},
  numpages = {6},
  year = {2013},
  month = {Mar},
  publisher = {American Physical Society},
}

@article{McDonnell2022subradiantedge,
  doi = {10.22331/q-2022-09-15-805},
  url = {https://doi.org/10.22331/q-2022-09-15-805},
  title = {Subradiant edge states in an atom chain with waveguide-mediated hopping},
  author = {McDonnell, Ciaran and Olmos, Beatriz},
  journal = {{Quantum}},
  issn = {2521-327X},
  publisher = {{Verein zur F{\"{o}}rderung des Open Access Publizierens in den Quantenwissenschaften}},
  volume = {6},
  pages = {805},
  month = sep,
  year = {2022}
}

@article{Ray_2020Light,
doi = {10.1088/1367-2630/ab8265},
url = {https://doi.org/10.1088/1367-2630/ab8265},
year = {2020},
month = {jun},
publisher = {IOP Publishing},
volume = {22},
number = {6},
pages = {062001},
author = {Ray, Tridib and Gupta, Ratnesh K and Gokhroo, Vandna and Everett, Jesse L and Nieddu, Thomas and Rajasree, Krishnapriya S and {Nic Chormaic}, Síle},
title = {Observation of the $^{87}${Rb} 5{S}$_{1/2}$ to 4{D}$_{3/2}$ electric quadrupole transition at 516.6~nm mediated via an optical nanofibre},
journal = {New Journal of Physics}
}

@article{413r-dn5p,
  title = {Ferromagnetic traps for quasicontinuous operation of optical nanofiber interfaces},
  author = {Liu, Ruijuan and Wu, Jinggu and Jiang, Yuan and Zhao, Yanting and Wu, Saijun},
  journal = {Phys. Rev. Appl.},
  volume = {24},
  issue = {3},
  pages = {034015},
  numpages = {17},
  year = {2025},
  month = {Sep},
  publisher = {American Physical Society},
  doi = {10.1103/413r-dn5p},
  url = {https://link.aps.org/doi/10.1103/413r-dn5p}
}

@article{Vylegzhanin_2025c,
doi = {10.1088/1367-2630/adf058},
url = {https://doi.org/10.1088/1367-2630/adf058},
year = {2025},
month = {jul},
publisher = {IOP Publishing},
volume = {27},
number = {7},
pages = {073203},
author = {Vylegzhanin, Alexey and Brown, Dylan J and Kornovan, Danil F and Brion, Etienne and {Nic Chormaic}, Síle},
title = {Towards a fictitious magnetic field trap for both ground and {R}ydberg state $^{87}${Rb} atoms via the evanescent field of an optical nanofiber},
journal = {New J. Phys.}
}

@article{PhysRevLett.132.113601,
  title = {Control and Entanglement of Individual {R}ydberg Atoms near a Nanoscale Device},
  author = {Ocola, Paloma L. and Dimitrova, Ivana and Grinkemeyer, Brandon and Guardado-Sanchez, Elmer and \DJ{}or\dj{}evi\ifmmode \acute{c}\else \'{c}\fi{}, Tamara and Samutpraphoot, Polnop and Vuleti\ifmmode \acute{c}\else \'{c}\fi{}, Vladan and Lukin, Mikhail D.},
  journal = {Phys. Rev. Lett.},
  volume = {132},
  issue = {11},
  pages = {113601},
  numpages = {6},
  year = {2024},
  month = {Mar},
  publisher = {American Physical Society},
  doi = {10.1103/PhysRevLett.132.113601},
  url = {https://link.aps.org/doi/10.1103/PhysRevLett.132.113601}
}

@article{89l1-dys4,
  title = {Magic-wavelength nanofiber-based two-color dipole trap with sub-$\ensuremath{\lambda}/2$ spacing},
  author = {Pache, Lucas and Cordier, Martin and Letellier, Hector and Schemmer, Max and Schneeweiss, Philipp and Volz, J\"urgen and Rauschenbeutel, Arno},
  journal = {Phys. Rev. A},
  volume = {112},
  issue = {1},
  pages = {L011701},
  numpages = {6},
  year = {2025},
  month = {Jul},
  publisher = {American Physical Society},
  doi = {10.1103/89l1-dys4},
  url = {https://link.aps.org/doi/10.1103/89l1-dys4}
}

@article{thompson2013coupling,
  title={Coupling a single trapped atom to a nanoscale optical cavity},
  author={Thompson, Jeffrey Douglas and Tiecke, TG and de Leon, Nathalie P and Feist, J and Akimov, AV and Gullans, M and Zibrov, Alexander S and Vuleti{\'c}, V and Lukin, Mikhail D},
  journal={Science},
  volume={340},
  number={6137},
  pages={1202--1205},
  year={2013},
  publisher={American Association for the Advancement of Science}
}

@article{dhordjevic2021entanglement,
  title={Entanglement transport and a nanophotonic interface for atoms in optical tweezers},
  author={{\DH}or{\dj}evi{\'c}, Tamara and Samutpraphoot, Polnop and Ocola, Paloma L and Bernien, Hannes and Grinkemeyer, Brandon and Dimitrova, Ivana and Vuleti{\'c}, Vladan and Lukin, Mikhail D},
  journal={Science},
  volume={373},
  number={6562},
  pages={1511--1514},
  year={2021},
  publisher={American Association for the Advancement of Science}
}

@article{corzo2019waveguide,
  title={Waveguide-coupled single collective excitation of atomic arrays},
  author={Corzo, Neil V and Raskop, J{\'e}r{\'e}my and Chandra, Aveek and Sheremet, Alexandra S and Gouraud, Baptiste and Laurat, Julien},
  journal={Nature},
  volume={566},
  number={7744},
  pages={359--362},
  year={2019},
  publisher={Nature Publishing Group UK London}
}

@article{berroir2025ultralow,
  title={Ultralow-power single-pass all-optical photon router},
  author={Berroir, J{\'e}r{\'e}my and Ray, Tridib and Urvoy, Alban and Laurat, Julien},
  journal={Optica},
  volume={12},
  number={8},
  pages={1250--1251},
  year={2025},
  publisher={Optica Publishing Group}
}

@article{Memory2015,
  title = {Demonstration of a Memory for Tightly Guided Light in an Optical Nanofiber},
  author = {Gouraud, B. and Maxein, D. and Nicolas, A. and Morin, O. and Laurat, J.},
  journal = {Phys. Rev. Lett.},
  volume = {114},
  issue = {18},
  pages = {180503},
  numpages = {5},
  year = {2015},
  month = {May},
  publisher = {American Physical Society},
}

@article{stourm2020spontaneous,
  title={Spontaneous emission and energy shifts of a {R}ydberg rubidium atom close to an optical nanofiber},
  author={Stourm, E and Lepers, Maxence and Robert, J and {Nic Chormaic}, S and M{\o}lmer, K and Brion, Etienne},
  journal={Phys. Rev. A},
  volume={101},
  number={5},
  pages={052508},
  year={2020},
  publisher={APS}
}

@article{stourm2023interaction,
  title={Interaction of two {R}ydberg atoms in the vicinity of an optical nanofibre},
  author={Stourm, Erwan and Lepers, Maxence and Robert, Jacques and {Nic Chormaic}, Sile and M{\o}lmer, Klaus and Brion, Etienne},
  journal={New J. Phys.},
  volume={25},
  number={2},
  pages={023022},
  year={2023},
  publisher={IOP Publishing}
}

@article{ocola2024control,
  title={Control and entanglement of individual {R}ydberg atoms near a nanoscale device},
  author={Ocola, Paloma L and Dimitrova, Ivana and Grinkemeyer, Brandon and Guardado-Sanchez, Elmer and {\DJ}or{\dj}evi{\'c}, Tamara and Samutpraphoot, Polnop and Vuleti{\'c}, Vladan and Lukin, Mikhail D},
  journal={Phys. Rev. Lett.},
  volume={132},
  number={11},
  pages={113601},
  year={2024},
  publisher={APS}
}

@article{sensing1999,
  title = {Using High {R}ydberg States as Electric Field Sensors},
  author = {Osterwalder, A. and Merkt, F.},
  journal = {Phys. Rev. Lett.},
  volume = {82},
  issue = {9},
  pages = {1831--1834},
  numpages = {0},
  year = {1999},
  month = {Mar},
  publisher = {American Physical Society},
}

@article{le2018enhancement,
  title={Enhancement of the quadrupole interaction of an atom with the guided light of an ultrathin optical fiber},
  author={Le Kien, Fam and Ray, Tridib and Nieddu, Thomas and Busch, Thomas and {Nic Chormaic}, S{\'\i}le},
  journal={Phys. Rev. A},
  volume={97},
  number={1},
  pages={013821},
  year={2018},
  publisher={APS}
}

@article{li2023super,
  title={Super low-frequency electric field measurement based on {R}ydberg atoms},
  author={Li, Ling and Jiao, Yuechun and Hu, Jinlian and Li, Huaqiang and Shi, Meng and Zhao, Jianming and Jia, Suotang},
  journal={Opt. Express},
  volume={31},
  number={18},
  pages={29228--29234},
  year={2023},
  publisher={Optica Publishing Group}
}

@article{bhavya2026interfacing,
  title={Interfacing an array of single atom tweezers with fiber guided mode},
  author={Bhavya, Puthanveettil and Iidawa, Kei and Morinaga, Makoto and Nayak, Kali P},
  journal={Appl. Phys. Express},
  volume={19},
  number={4},
  pages={042001},
  year={2026},
  publisher={IOP Publishing}
}

@article{pache2026lambda,
title={${\Lambda}$-enhanced gray-molasses loading and {EIT} cooling of neutral atoms in nanophotonic traps},
author={Pache, Lucas and Glicenstein, Antoine and Schneeweiss, Philipp and Volz, J{\"u}rgen and Rauschenbeutel, Arno and Pennetta, Riccardo},
journal={arXiv:2605.13387},
year={2026},
}

@article{Aoki2015,
  title = {Strong Coupling between a Trapped Single Atom and an All-Fiber Cavity},
  author = {Kato, Shinya and Aoki, Takao},
  journal = {Phys. Rev. Lett.},
  volume = {115},
  issue = {9},
  pages = {093603},
  numpages = {5},
  year = {2015},
  month = {Aug},
  publisher = {American Physical Society},
}

@article{takahata2026fiber,
  title={Fiber-optic quantum interface with an array of more than 100 individually addressable atoms on an optical nanofiber},
  author={Takahata, Mitsuyoshi and Keloth, Jameesh and Yamamoto, Takashi and Harada, Ken Ichi and Miki, Shigehito and Aoki, Takao},
  journal={arXiv:2603.21812},
  year={2026}
}

@article{vylegzhanin2023excitation,
title={Excitation of $^{87}${R}b {R}ydberg atoms to $n${S} and $n${D} states ($n\leq 68$) via an optical nanofiber},
author={Vylegzhanin, Alexey and Brown, Dylan J and Raj, Aswathy and Kornovan, Danil F and Everett, Jesse L and Brion, Etienne and Robert, Jacques and {Nic Chormaic}, S{\'\i}le},
journal={Optica Quantum},
volume={1},
number={1},
pages={6--13},
year={2023},
publisher={Optica Publishing Group},
doi= {https://doi.org/10.1364/OPTICAQ.498414}
}

@article{self-org2016,
  title = {Self-organization of atoms coupled to a chiral reservoir},
  author = {Eldredge, Zachary and Solano, Pablo and Chang, Darrick and Gorshkov, Alexey V.},
  journal = {Phys. Rev. A},
  volume = {94},
  issue = {5},
  pages = {053855},
  numpages = {13},
  year = {2016},
  month = {Nov},
  publisher = {American Physical Society},
  doi = {10.1103/PhysRevA.94.053855},
  url = {https://link.aps.org/doi/10.1103/PhysRevA.94.053855}
}

@inproceedings{NicChormaic:06,
author = {S\'{i}le {Nic Chormaic} and Michael Morrissey and Kieran Deasy and Thejesh B. Nagabhushan and Jonathan Ward and Brian Shortt},
booktitle = {Frontiers in Optics},
journal = {Frontiers in Optics},
pages = {LTuK3},
publisher = {Optica Publishing Group},
title = {Evanescent Field Atom Optics Using Micro-Tapered Fibers},
year = {2006},
url = {https://opg.optica.org/abstract.cfm?URI=LS-2006-LTuK3},
doi = {10.1364/LS.2006.LTuK3},
}

@article{Warken:07,
author = {F. Warken and E. Vetsch and D. Meschede and M. Sokolowski and A. Rauschenbeutel},
journal = {Opt. Express},
number = {19},
pages = {11952--11958},
publisher = {Optica Publishing Group},
title = {Ultra-sensitive surface absorption spectroscopy using sub-wavelength diameter optical fibers},
volume = {15},
month = {Sep},
year = {2007},
url = {https://opg.optica.org/oe/abstract.cfm?URI=oe-15-19-11952},
doi = {10.1364/OE.15.011952},
}

@article{Lamsal:19,
author = {H. P. Lamsal and J. D. Franson and T. B. Pittman},
journal = {Appl. Opt.},
number = {24},
pages = {6470--6473},
publisher = {Optica Publishing Group},
title = {Transmission characteristics of optical nanofibers in metastable xenon},
volume = {58},
month = {Aug},
year = {2019},
url = {https://opg.optica.org/ao/abstract.cfm?URI=ao-58-24-6470},
doi = {10.1364/AO.58.006470},
}

@article{rajasree2020generation,
title={Generation of cold {R}ydberg atoms at submicron distances from an optical nanofiber},
author={Rajasree, Krishnapriya Subramonian and Ray, Tridib and Karlsson, Kristoffer and Everett, Jesse L and {Nic Chormaic}, S{\'\i}le},
journal={Phys. Rev. Res.},
volume={2},
pages={012038},
year={2020},
publisher={APS},
doi = {https://doi.org/10.1103/PhysRevResearch.2.012038}
}

@article{viteau2011rydberg,
  title={{R}ydberg spectroscopy of a {R}b {M}{O}{T} in the presence of applied or ion created electric fields},
  author={Viteau, M and Radogostowicz, J and Bason, Mark George and Malossi, Nicola and Ciampini, Donatella and Morsch, O and Arimondo, Ennio},
  journal={Opt. Express},
  volume={19},
  number={7},
  pages={6007--6019},
  year={2011},
  publisher={Optical Society of America}
}

@article{grabowski2006high,
  title={High resolution {R}ydberg spectroscopy of ultracold rubidium atoms},
  author={Grabowski, Axel and Heidemann, Rolf and L{\"o}w, Robert and Stuhler, J{\"u}rgen and Pfau, Tilman},
  journal={Fortschr. Phys.},
  volume={54},
  number={8-10},
  pages={765--775},
  year={2006},
  publisher={Wiley Online Library}
}

@article{rajasree20211,
  title={1.6 {GHz} frequency scanning of a 482 nm laser stabilized using electromagnetically induced transparency},
  author={Rajasree, Krishnapriya Subramonian and Karlsson, Kristoffer and Ray, Tridib and {Nic Chormaic}, S{\'\i}le},
  journal={IEEE Photonics Technol. Lett.},
  volume={33},
  number={15},
  pages={780--783},
  year={2021},
  publisher={IEEE}
}

@article{tkachenko2019polarisation,
  title={Polarisation control for optical nanofibres by imaging through a single lens},
  author={Tkachenko, Georgiy and Lei, Fuchuan and {Nic Chormaic}, S{\'\i}le},
  journal={J. Opt.},
  volume={21},
  number={12},
  pages={125604},
  year={2019},
  publisher={IOP Publishing}
}

@article{bai2019single,
  title={Single-photon {R}ydberg excitation and trap-loss spectroscopy of cold cesium atoms in a magneto-optical trap by using of a 319-nm ultraviolet laser system},
  author={Bai, Jiandong and Liu, Shuo and Wang, Jieying and He, Jun and Wang, Junmin},
  journal={IEEE J. Sel. Top. Quantum Electron.},
  volume={26},
  number={3},
  pages={1--6},
  year={2019},
  publisher={IEEE}
}

@article{sunami2025scalable,
  title={Scalable Networking of Neutral-Atom Qubits: Nanofiber-Based Approach for Multiprocessor Fault-Tolerant Quantum Computers},
  author={Sunami, Shinichi and Tamiya, Shiro and Inoue, Ryotaro and Yamasaki, Hayata and Goban, Akihisa},
  journal={PRX Quantum},
  volume={6},
  number={1},
  pages={010101},
  year={2025},
  publisher={APS}
}

@article{saffman2016quantum,
  title={Quantum computing with atomic qubits and {R}ydberg interactions: progress and challenges},
  author={Saffman, Mark},
  journal={J. Phys. B: At. Mol. Opt. Phys.},
  volume={49},
  number={20},
  pages={202001},
  year={2016},
  publisher={IOP Publishing}
}

@article{adams2019rydberg,
  title={{R}ydberg atom quantum technologies},
  author={Adams, Charles S and Pritchard, Jonathan D and Shaffer, James P},
  journal={J. Phys. B: At. Mol. Opt. Phys.},
  volume={53},
  number={1},
  pages={012002},
  year={2019},
  publisher={IOP Publishing}
}

@article{ding2022enhanced,
  title={Enhanced metrology at the critical point of a many-body {R}ydberg atomic system},
  author={Ding, Dong-Sheng and Liu, Zong-Kai and Shi, Bao-Sen and Guo, Guang-Can and M{\o}lmer, Klaus and Adams, Charles S},
  journal={Nat. Phys.},
  volume={18},
  number={12},
  pages={1447--1452},
  year={2022},
  publisher={Nature Publishing Group UK London}
}

@article{gupta2022machine,
  title={Machine learner optimization of optical nanofiber-based dipole traps},
  author={Gupta, Ratnesh K and Everett, Jesse L and Tranter, Aaron D and Henke, Ren{\'e} and Gokhroo, Vandna and Lam, Ping Koy and {Nic Chormaic}, S{\'\i}le},
  journal={AVS Quantum Sci.},
  volume={4},
  number={2},
  year={2022},
  publisher={AIP Publishing}
}

@article{vsibalic2017arc,
  title={{ARC}: An open-source library for calculating properties of alkali {R}ydberg atoms},
  author={{\v{S}}ibali{\'c}, Nikola and Pritchard, Jonathan D and Adams, Charles S and Weatherill, Kevin J},
  journal={Comput. Phys. Commun.},
  volume={220},
  pages={319--331},
  year={2017},
  publisher={Elsevier}
}

@article{lindblad1976generators,
  title={On the generators of quantum dynamical semigroups},
  author={Lindblad, Goran},
  journal={Commun. Math. Phys.},
  volume={48},
  number={2},
  pages={119--130},
  year={1976},
  publisher={Springer}
}

@article{vetsch2010optical,
  title={Optical interface created by laser-cooled atoms trapped in the evanescent field surrounding an optical nanofiber},
  author={Vetsch, E and Reitz, D and Sagu{\'e}, G and Schmidt, R and Dawkins, ST and Rauschenbeutel, Arno},
  journal={Phys. Rev. Lett.},
  volume={104},
  number={20},
  pages={203603},
  year={2010},
  publisher={APS}
}

@article{LeKien2004,
  title = {Atom trap and waveguide using a two-color evanescent light field around a subwavelength-diameter optical fiber},
  author = {Le Kien, Fam and Balykin, V. I. and Hakuta, K.},
  journal = {Phys. Rev. A},
  volume = {70},
  issue = {6},
  pages = {063403},
  numpages = {9},
  year = {2004},
  month = {Dec},
  publisher = {American Physical Society},
  doi = {10.1103/PhysRevA.70.063403},
  
}

@phdthesis{Weller2019,
  author  = {Daniel Weller},
  title   = {{Thermal {R}ydberg Spectroscopy and Plasma}},
  school  = {University of Stuttgart},
  year    = {2019}
}

@article{abel2011electrometry,
  title={Electrometry near a dielectric surface using {R}ydberg electromagnetically induced transparency},
  author={Abel, R P and Carr, C and Krohn, U and Adams, C S},
  journal={Phys. Rev. A},
  volume={84},
  number={2},
  pages={023408},
  year={2011},
  publisher={APS}
}

@article{neufeld2011probing,
  title={Probing stray surface electric patch fields using {R}ydberg atoms},
  author={Neufeld, DD and Pu, Y and Dunning, FB},
  journal={Nucl. Instrum. Methods Phys. Res. Sect. B},
  volume={269},
  number={11},
  pages={1288--1291},
  year={2011},
  publisher={Elsevier}
}

@article{nayak2019real,
  title={Real-time observation of single atoms trapped and interfaced to a nanofiber cavity},
  author={Nayak, Kali P and Wang, Jie and Keloth, Jameesh},
  journal={Phys. Rev. Lett.},
  volume={123},
  number={21},
  pages={213602},
  year={2019},
  publisher={APS}
}

@article{skljarow2022purcell,
  title={Purcell-enhanced dipolar interactions in nanostructures},
  author={Skljarow, Artur and K{\"u}bler, Harald and Adams, Charles S and Pfau, Tilman and L{\"o}w, Robert and Alaeian, Hadiseh},
  journal={Phys. Rev. Res.},
  volume={4},
  number={2},
  pages={023073},
  year={2022},
  publisher={APS}
}

@article{wave_guide,
  title = {Waveguide quantum electrodynamics: Collective radiance and photon-photon correlations},
  author = {Sheremet, Alexandra S. and Petrov, Mihail I. and Iorsh, Ivan V. and Poshakinskiy, Alexander V. and Poddubny, Alexander N.},
  journal = {Rev. Mod. Phys.},
  volume = {95},
  issue = {1},
  pages = {015002},
  numpages = {59},
  year = {2023},
  month = {Mar},
  publisher = {American Physical Society},
}

@article{le2004field,
  title={Field intensity distributions and polarization orientations in a vacuum-clad subwavelength-diameter optical fiber},
  author={Le Kien, Fam and Liang, JQ and Hakuta, K and Balykin, VI},
  journal={Opt. Commun.},
  volume={242},
  number={4-6},
  pages={445--455},
  year={2004},
  publisher={Elsevier}
}

@article{negative_electron_affinity,
  title = {Electric field cancellation on quartz by {R}b adsorbate-induced negative electron affinity},
  author = {Sedlacek, J. A. and Kim, E. and Rittenhouse, S. T. and Weck, P. F. and Sadeghpour, H. R. and Shaffer, J. P.},
  journal = {Phys. Rev. Lett.},
  volume = {116},
  issue = {13},
  pages = {133201},
  numpages = {7},
  year = {2016},
  month = {Mar},
  publisher = {American Physical Society},
}

@article{chan2014adsorbate,
  title={Adsorbate electric fields on a cryogenic atom chip},
  author={Chan, KS and Siercke, M and Hufnagel, C and Dumke, R},
  journal={Phys. Rev. Lett.},
  volume={112},
  number={2},
  pages={026101},
  year={2014},
  publisher={APS}
}

@article{hattermann2012detrimental,
  title={Detrimental adsorbate fields in experiments with cold {R}ydberg gases near surfaces},
  author={Hattermann, H and Mack, M and Karlewski, F and Jessen, F and Cano, D and Fort{\'a}gh, J},
  journal={Phys. Rev. A},
  volume={86},
  number={2},
  pages={022511},
  year={2012},
  publisher={APS}
}

@article{epple2017effect,
  title={Effect of stray fields on {R}ydberg states in hollow-core {P}{C}{F} probed by higher-order modes},
  author={Epple, Georg and Joly, Nicolas Y and Euser, Tijmen G and St. J. Russell, P and L{\"o}w, Robert},
  journal={Opt. Lett.},
  volume={42},
  number={17},
  pages={3271--3274},
  year={2017},
  publisher={Optical Society of America}
}

@article{mcguirk2004alkali,
  title={Alkali-metal adsorbate polarization on conducting and insulating surfaces probed with {B}ose-{E}instein condensates},
  author={McGuirk, JM and Harber, DM and Obrecht, John M and Cornell, Eric A},
  journal={Phys. Rev. A},
  volume={69},
  number={6},
  pages={062905},
  year={2004},
  publisher={APS}
}

@article{sedlacek2016electric,
  title={Electric field cancellation on quartz by {R}b adsorbate-induced negative electron affinity},
  author={Sedlacek, JA and Kim, Eunja and Rittenhouse, ST and Weck, PF and Sadeghpour, HR and Shaffer, JP},
  journal={Phys. Rev. Lett.},
  volume={116},
  number={13},
  pages={133201},
  year={2016},
  publisher={APS}
}

@article{tauschinsky2010spatially,
  title={Spatially resolved excitation of {R}ydberg atoms and surface effects on an atom chip},
  author={Tauschinsky, Atreju and Thijssen, Rutger M T and Whitlock, S and van Linden van den Heuvell, H B and Spreeuw, R J C},
  journal={Phys. Rev. A},
  volume={81},
  number={6},
  pages={063411},
  year={2010},
  publisher={APS}
}

@article{weller2016charge,
  title={Charge-induced optical bistability in thermal {R}ydberg vapor},
  author={Weller, Daniel and Urvoy, Alban and Rico, Andy and L{\"o}w, Robert and K{\"u}bler, Harald},
  journal={Phys. Rev. A},
  volume={94},
  number={6},
  pages={063820},
  year={2016},
  publisher={APS}
}

@article{weller2019interplay,
  title={Interplay between thermal {R}ydberg gases and plasmas},
  author={Weller, Daniel and Shaffer, James P and Pfau, Tilman and L{\"o}w, Robert and K{\"u}bler, Harald},
  journal={Phys. Rev. A},
  volume={99},
  number={4},
  pages={043418},
  year={2019},
  publisher={APS}
}

@article{sedlacek2012microwave,
  title={Microwave electrometry with {R}ydberg atoms in a vapour cell using bright atomic resonances},
  author={Sedlacek, Jonathon A and Schwettmann, Arne and K{\"u}bler, Harald and L{\"o}w, Robert and Pfau, Tilman and Shaffer, James P},
  journal={Nat. Phys.},
  volume={8},
  number={11},
  pages={819--824},
  year={2012},
  publisher={Nature Publishing Group UK London}
}

@article{holloway2022rydberg,
  title={{R}ydberg atom-based field sensing enhancement using a split-ring resonator},
  author={Holloway, Christopher L and Prajapati, Nikunjkumar and Artusio-Glimpse, Alexandra B and Berweger, Samuel and Simons, Matthew T and Kasahara, Yoshiaki and Alu, Andrea and Ziolkowski, Richard W},
  journal={Appl. Phys. Lett.},
  volume={120},
  number={20},
  year={2022},
  pages={204001},
  publisher={AIP Publishing}
}

@article{vylegzhanin2026light,
  title={Light-induced, fictitious magnetic trapping of cold alkali-metal atoms using an optical-tweezer--nanofiber hybrid platform},
  author={Vylegzhanin, Alexey and Brown, Dylan J and Abdrakhmanov, Sergei and {Nic Chormaic}, S{\'\i}le },
  journal={Phys. Rev. A},
  volume={113},
  number={2},
  pages={023111},
  year={2026},
  publisher={APS}
}

@article{nedeljkovic2005ionization,
  title={Ionization distances of {R}ydberg atoms approaching solid surfaces in the presence of weak electric fields},
  author={Nedeljkovi{\'c}, NN and Nedeljkovi{\'c}, Lj D},
  journal={Phys. Rev. A},
  volume={72},
  number={3},
  pages={032901},
  year={2005},
  publisher={APS}
}

@article{kohlhoff2016interaction,
  title={Interaction of {R}ydberg atoms with surfaces: Using surface ionisation as a probe for surface analysis},
  author={Kohlhoff, Mike W},
  journal={Eur. Phys. J. Spec. Top. },
  volume={225},
  number={15},
  pages={3061--3085},
  year={2016},
  publisher={Springer}
}

@article{zhang2026microwave,
  title={Microwave electrometry with quantum-limited resolutions in a {R}ydberg-atom array},
  author={Zhang, Yao-Wen and Xiang, De-Sheng and Liao, Ren and Liu, Hao-Xiang and Xu, Biao and Zhou, Peng and Zhou, Yijia and Zhang, Kuan and Li, Lin},
  journal={Phys. Rev. Lett.},
  volume={136},
  number={11},
  pages={110802},
  year={2026},
  publisher={APS}
}

@article{meyer2020assessment,
  title={Assessment of {R}ydberg atoms for wideband electric field sensing},
  author={Meyer, David H and Castillo, Zachary A and Cox, Kevin C and Kunz, Paul D},
  journal={J. Phys. B: At. Mol. Opt. Phys.},
  volume={53},
  number={3},
  pages={034001},
  year={2020},
  publisher={IOP Publishing}
}

@article{gallagher1988rydberg,
  title={{R}ydberg atoms},
  author={Gallagher, Thomas F},
  journal={Rep. Prog. Phys.},
  volume={51},
  number={2},
  pages={143--188},
  year={1988}
}

@article{harter2012single,
  title={Single ion as a three-body reaction center in an ultracold atomic gas},
  author={H{\"a}rter, Arne and Kr{\"u}kow, Artjom and Brunner, Andreas and Schnitzler, Wolfgang and Schmid, Stefan and Denschlag, Johannes Hecker},
  journal={Phys. Rev. Lett.},
  volume={109},
  number={12},
  pages={123201},
  year={2012},
  publisher={APS}
}

@article{pennetta2025hybrid,
  title={Hybrid nanophotonic trap for cold atoms using surface forces and a blue-detuned evanescent field},
  author={Pennetta, Riccardo and Glicenstein, Antoine and Pache, Lucas and Schneeweiss, Philipp and Volz, J{\"u}rgen and Rauschenbeutel, Arno},
  journal={Nat. Photon.},
  year={2026},
  pages={1-6}
}

@article{Wuttke2013,
    title={Optically active mechanical modes of tapered optical fibers},
    author={Wuttke, Chrisitan and Cole, Garrett D. and Rauschenbeutel, Arno},
    journal={Phys. Rev. A},
    year={2013},
    volume={88},
    pages={061801(R)}
}

@article{Tebbenjohanns2023,
    title={Feedback cooling the fundamental torsional mechanical mode of a tapered optical fiber to 30 m{K}},
    author={Tebbenjohanns, Felix and Jia, Jihao and Antesberger, Michael and Prasad, Adarsh Shankar and Pucher, Sebastian and Rauschenbeutel, Arno and Volz, J{\"u}rgen and Schneeweiss, Philipp},
    journal={Phys. Rev. A},
    year={2023},
    volume={108(3)},
    pages={L031101}
}

\end{document}